\documentstyle[aps,preprint,tighten,eqsecnum]{revtex}
\begin{document}
\hyphenation{Nijmegen}
\hyphenation{Rijken}
\draft
\preprint{\vbox{. \hfill TRI-PP-96-61}}

\title{Meson-baryon coupling constants from a chiral-invariant SU(3)
       Lagrangian and application to $\protect\bbox{N\!N}$ scattering}
\author{V.G.J.\ Stoks\thanks{Electronic address: stoks@alph02.triumf.ca}}
\address{TRIUMF, 4004 Wesbrook Mall, Vancouver,
         British Columbia, Canada V6T 2A3}
\author{Th.A.\ Rijken}
\address{Institute for Theoretical Physics, University of Nijmegen,
         Nijmegen, The Netherlands}
\date{submitted to Nucl.\ Phys.\ A; manuscript NUCPHA 1336}
\maketitle

\begin{abstract}
We present a chiral-invariant meson-baryon Lagrangian which describes
the interactions of the baryon octet with the lowest-mass meson nonets.
The nonlinear realization of the chiral symmetry generates
pair-meson interaction vertices. The corresponding pair-meson
coupling constants can all be expressed in terms of the
meson-nucleon-nucleon pseudovector, scalar, and vector coupling
constants, and their corresponding $F/(F+D)$ ratios, and for which
empirical estimates are given. We show that it is possible to
construct an $N\!N$ potential of reasonable quality satisfying these
theoretical and empirical constraints. 
\end{abstract}
%\narrowtext
%\begin{multicols}

\section{Introduction}
\label{sec:intro}
The construction of $N\!N$ potentials has a long history, and a large
number of different models have appeared in the literature. After a
comparison of these $N\!N$ potentials by confronting them first with
the $pp$ scattering data~\cite{Sto93a} and later also with the $np$
scattering data~\cite{Sto95}, it was found that all existing models
do a rather poor job (some worse than others) in properly describing
these data. This is rather disturbing, since most of these models are
used in many-body calculations and $pp$ bremsstrahlung calculations
to draw certain conclusions on three-body forces, relativistic effects,
off-shell effects, etc. And so one can ask the question whether it
is at all valid to conclude anything from a many-body calculation
if the applied $N\!N$ potential cannot even adequately describe the
two-nucleon data.

This uncertainty has been recently cleared up with the construction
of a number of high-quality $N\!N$ potential models~\cite{Sto94a,Wir95}.
With high-quality we mean that these models describe the $N\!N$
scattering data with the almost optimal $\chi^{2}/N_{\rm data}\approx1$.
Although based on very different functional forms, these potentials
give remarkably similar results in many-body~\cite{Fri93,Zhe94,Glo96}
and $pp$ bremsstrahlung~\cite{Ede96} calculations. One `disadvantage'
of these new potential models, however, is that they are largely
phenomenological and only explicitly include the one-pion exchange.
Our next task, therefore, is to try and construct a high-quality
potential model which is less phenomenological in that it explicitly
contains the heavier-meson exchanges (with coupling constants which
can be compared with empirical data) and which, possibly, is consistent
with the symmetries of QCD.

Recently, Ord\'o\~nez, Ray, and van Kolck~\cite{Ord96} presented a
nucleon-nucleon ($N\!N$) potential based on an effective chiral
Lagrangian of pions, nucleons, and $\Delta$ isobars. Using 26 free
parameters, the agreement with the experimental scattering data was
found to be satisfactory up to lab energies of about 100 MeV.
An extension to higher energies and a further improvement in the
description of the data would require an expansion to higher orders in
chiral perturbation theory, making the model much more complicated and
introducing many new parameters. Hence, the authors conclude that it
is not practical for potential models derived from effective chiral
Lagrangians to compete with more phenomenological approaches.

In this paper we would like to advocate an alternative approach.
We want to investigate whether it is possible to construct a potential
model which not only gives a satisfactory description of the scattering
data up to $\sim$350 MeV using a limited number of free parameters, but
which also retains the salient features of chiral symmetry.
For that purpose we here do not integrate out all mesons other than the
pion~\cite{Wei90}, but rather adopt the successful approach used in
one-boson-exchange models and keep all lowest-lying mesons with masses
lower than 1 GeV, say. The $N\!N$ potential model is then obtained by
evaluating the standard one-boson-exchange contributions involving these
mesons, but now including the contributions of the box and crossed-box
two-meson diagrams~\cite{Rij96a} and of the pair-meson diagrams where at
least one of the nucleon lines contains a pair-meson vertex~\cite{Rij96b}.
The potential contributions are calculated up to order $1/M^{2}$ in
the nucleon mass as is customary in the conventional one-boson-exchange
approaches, and so we want to stress that here we are {\it not\/} doing
chiral perturbation theory. Here we only use chiral symmetry to
generate the interaction Lagrangians and to find constraints for
the associated coupling constants. However, as we will see, the
pair-meson diagrams arise as a direct consequence of chiral symmetry.
The box and crossed-box diagrams then also have to be included because
they are of the same order in the number of exchanged mesons as the
pair-meson diagrams.
The pair-meson interactions can be viewed~\cite{Wei90} as the result of
integrating out the heavy-meson (masses larger than 1 GeV, say) and
resonance (e.g., $\Delta$, $N^{\ast}$, $Y^{\ast}$) degrees of freedom
in the two-meson-exchange processes.
Also, according to ``duality''~\cite{Dol68}, the resonance contributions
to the various meson-nucleon amplitudes can be described approximately
by heavy-meson exchanges. Treating the heavy-meson propagators as
constants, which should be adequate at low energies, then leads
directly to pair-meson exchanges.

In Refs.~\cite{Rij96a,Rij96b} we already showed that the inclusion of
the two-meson (box, crossed-box, and pair) contributions provides a
substantial improvement in the description of the scattering data as
compared to a potential containing only the standard one-boson exchanges.
Although originally~\cite{Rij93,Sto94b} the meson-pair interaction
Lagrangians were taken to be purely phenomenological, we later found
that the pair-meson coupling constants could all be fixed using
experimental input and chiral-symmetry constraints. In particular,
in Ref.~\cite{Rij96b} the estimates for the pair-meson coupling
constants were based on the linear $\sigma$ model~\cite{Gel60}.
In order to appreciate this result, it should be realized that, by
fixing the pair-meson coupling constants in this way, this improvement
could be obtained without the introduction of any new parameters.

This remarkable result encourages us to go beyond the $N\!N$ model and
to investigate whether a similar approach will also be fruitful in the
construction of an extended hyperon-nucleon ($Y\!N$) potential.
An important motivation for the development of an extended $Y\!N$
model is provided by the study of hypernuclei using one-boson-exchange
models~\cite{Yam90,Car90}. The problem with the construction of a
$Y\!N$ potential is that there are only a few experimental data
available, and these data are rather old and not very accurate.
At present it is very difficult (if not impossible) to determine all
the free parameters from the scattering data alone.
A reduction in the number of free parameters is obtained by first
fixing the parameters which also play a role in $N\!N$ scattering
(and which are easier to determine, since there are many accurate
$N\!N$ scattering data), and then using SU(3) symmetry to fix the
coupling constants in the $Y\!N$ potential. This approach has been
used successfully in the various hard-core~\cite{Nag77} and
soft-core~\cite{Mae89} $Y\!N$ one-boson-exchange models of the Nijmegen
group. The J\"ulich models~\cite{Hol89,Reu94} even impose an SU(6)
symmetry. As a further reduction in the number of parameters, it
would be convenient if we were able also to fix all (or at least most)
single-meson coupling constants at their empirical values.

As a first step, in this paper we will construct a chiral-invariant
Lagrangian for the meson-baryon sector. Rather than extending the
linear $\sigma$ model (the model we used in \cite{Rij96b}) to the
strange sector, we here find it more convenient to explore the
nonlinear realization of the spontaneously broken chiral symmetry.
This allows us to express all pair-meson coupling constants in terms
of the single-meson vertex parameters. Using symmetry arguments and
further experimental data we give estimates for these single-meson
vertex parameters.
We then have two options for constructing a potential model: (i) we can
either strictly impose all the constraints on the coupling constants
and investigate whether it is at all possible to impose chiral
symmetry in a potential model or, (ii) we can relax at least some of the
constraints and, hence, optimize the description of the scattering data.
As an application we will here only explore the first option and show
that a fully constrained $N\!N$ model indeed allows for a very
satisfactory description of the data. This is a remarkable result
because, until now, in potential models it has never been possible to
fix more than only a few coupling constants at their empirical values.
The second option reflects that the imposed symmetries need not be
exactly true, a fact which can be useful when fine tuning the model to
the scattering data. This option will be left for the future.

The outline of the paper is as follows. In Sec.~\ref{sec:content} we
list the baryons and mesons of the model. Section~\ref{sec:su2} gives
a review of the linear $\sigma$ model~\cite{Gel60} and the nonlinear
realization of chiral symmetry as introduced by Weinberg~\cite{Wei67}.
We briefly indicate how the model can be made to agree with experimental
data. In Sec.~\ref{sec:su3} we extend the nonlinear realization to
SU(3) to obtain a chiral-invariant meson-baryon Lagrangian. This SU(3)
Lagrangian describes the meson-baryon interactions of the scalar,
pseudoscalar, vector, and axial-vector meson nonets with the baryon
octet.
Section~\ref{sec:single-meson} then lists estimates for the various
single-meson vertex parameters (singlet and octet coupling constants,
mixing angles, and $F/(F+D)$ ratios), using theoretical and experimental
input. The coupling constants for the double-meson vertices can all
be expressed in terms of these single-meson vertex parameters;
this will be discussed in Sec.~\ref{sec:double-meson}. Unfortunately,
we have not been able to construct a potential model that satisfies all
these constraints. Therefore, in Sec.~\ref{subsec:extension} we briefly
indicate how the interaction Lagrangian can be extended by introducing
new free parameters without abandoning the empirical constraints on the
single-meson coupling constants. This allows for much more flexibility
in actually imposing these constraints in a baryon-baryon potential model.
As an application, in Sec.~\ref{sec:application} we show that with this
extension it is indeed possible to construct an $N\!N$ potential that
satisfies the constraints of chiral symmetry and which gives a very
satisfactory description of the $N\!N$ scattering data up to 350 MeV.

\section{Meson and baryon content}
\label{sec:content}
In the following, the baryons are the members of the SU(3) octet:
$N$(940), $\Lambda$(1115), $\Sigma$(1192), and $\Xi$(1318). The mesons
consist of the standard pseudoscalar nonet ($\pi$, $\eta$, $\eta'$, $K$)
and vector nonet ($\rho$, $\omega$, $\phi$, $K^{\ast}$). The physical
isoscalar mesons in these nonets come about as an admixture of the
pure octet and pure singlet isoscalar states. That is, the physical
$\eta$ and $\eta'$ are admixtures of the pure octet $\eta_{8}$ with
the pure singlet $\eta_{0}$, and the physical $\omega$ and $\phi$
are admixtures of the pure octet $\omega_{8}$ with the pure singlet
$\omega_{0}$.

In the SU(3)$\times$SU(3) chiral-invariant model, the extension of the
global chiral symmetry to a local chiral symmetry introduces two octets
of gauge fields, which can be re-expressed in terms of one octet of
vector and one octet of axial-vector fields (see below). It seems
natural to identify the octet of vector gauge fields with the octet
of vector mesons. The singlet (which is not a gauge field) is then
added to complete the nonet.
Similarly, the octet of axial-vector fields can be identified with the
octet of axial-vector mesons. Adding the singlet and mixing the pure
octet and pure singlet isoscalar states is then assumed to give the
physical nonet of axial-vector mesons [$a_{1}(1260)$, $f_{1}(1285)$,
$f_{1}(1420)$, and $K_{1}(1270)$].
In this scenario, the axial-vector mesons are a necessary ingredient
in establishing the local symmetry, but one can argue that their role
in low-energy baryon-baryon potential models is likely to be negligible.
Their masses are well above 1 GeV, and so the corresponding short-range
potentials are probably already effectively included by form factors at
the meson-baryon vertices.
For alternative approaches to include vector and axial-vector mesons
in a chiral-invariant way, see Refs.~\cite{Eck89,Jen95}.

The existence of the $0^{++}$ scalar nonet (octet plus singlet) is
still controversial. The Particle Data Group~\cite{PDG96} lists an
isovector state $a_{0}(980)$ and an isoscalar state $f_{0}(980)$, while
there appears to be evidence for a strange isodoublet of scalar mesons,
denoted by $\kappa(887)$~\cite{Sve92a}. These would make up the octet.
There is also evidence for a broad isoscalar $0^{++}(760)$
state~\cite{Sve92b}, which could be the scalar singlet.
Alternatively, in analogy with the pseudoscalar and vector nonets,
it is likely that also the $f_{0}(980)$ and $\varepsilon(760)$ are
admixtures of the pure octet and pure singlet isoscalar states.
The baryon-baryon potential due to the exchange of a broad meson
can be approximated by the sum of two potentials where each potential
is due to the exchange of a stable (effective) meson~\cite{Sch71}.
Due to the large width of the $\varepsilon(760)$, the low-mass pole
in this two-pole approximation is rather small ($\sim$500 MeV), and
hence would represent a possible candidate for the low-mass $\sigma$
meson which is a necessary ingredient in baryon-baryon potential
models, and which is the ingredient of the linear $\sigma$ model.
It is beyond the scope of this paper to discuss whether or not a
nonet of scalar mesons with masses below 1 GeV really exists,
how its existence can be explained or disclaimed in a quark model,
or whether a scalar one-boson exchange is nothing else but an effective
representation of correlated two-pion exchange.
Here we only mention that already some time ago Jaffe~\cite{Jaf77}
presented a quark-bag model calculation of $\bar{q}^{2}q^{2}$ mesons
where the members and decay modes of the lowest multiplet fit in
remarkably well with what is observed for the scalar nonet discussed
above. More recently, T\"ornqvist~\cite{Tor95} has calculated the
properties of a distorted $\bar{q}q$ nonet in a unitarized quark model,
which also identifies the above scalar nonet reasonably well (except
for the $\kappa(887)$ state). Finally, after an absence of almost 20
years, the latest edition of the Particle Data Group~\cite{PDG96} now
tentatively lists a very broad resonance as $f_{0}$(400--1200) with a
width of 600--1000 MeV. Here, we will stick to $\varepsilon(760)$ and
assume a width of $\Gamma_{\varepsilon}\approx800$ MeV.

\section{SU(2) chiral symmetry}
\label{sec:su2}
Although the SU(2) case has been discussed extensively in the
literature, we believe it is still instructive to first review the
SU(2) case in some detail before we turn to the SU(3) case.
In the SU(2) case the results can be easily expressed in terms of
the physical fields, which makes it easier to see what is going on.
In the SU(3) case the expressions are much more involved.

The classic example for a model with chiral SU(2) symmetry is the
linear $\sigma$ model~\cite{Gel60}. The linear $\sigma$ model contains
an isotriplet of pseudoscalars, $\bbox{\pi}$, and an isosinglet scalar,
$\sigma$, which can be grouped into
\begin{equation}
    \Sigma=\sigma+i\bbox{\tau}\!\cdot\!\bbox{\pi},
\end{equation}
and which transforms under global SU(2)$_{L}$$\times$SU(2)$_{R}$ as
\begin{equation}
   \Sigma \rightarrow  L \Sigma R^{\dagger}.  \label{Sigmatrans}
\end{equation}
The nucleon field $\psi$ has left and right components,
$\psi_{L,R}={\textstyle\frac{1}{2}}(1\mp\gamma_{5})\psi$,
transforming as
\begin{equation}
  \psi_{L}\rightarrow L\psi_{L}, \ \ \ \psi_{R}\rightarrow R\psi_{R}.
\end{equation}
Given these transformation properties, we can construct the
chiral-invariant Lagrangian ${\cal L}={\cal L}_{\psi}+{\cal L}_{\Sigma}
+{\cal L}_{I}$, where
\begin{eqnarray}
  {\cal L}_{\psi} &=&
     \overline{\psi}_{L}i\gamma^{\mu}\partial_{\mu}\psi_{L}
    +\overline{\psi}_{R}i\gamma^{\mu}\partial_{\mu}\psi_{R}, \nonumber\\
  {\cal L}_{\Sigma} &=& {\textstyle\frac{1}{4}}{\rm Tr}
     (\partial^{\mu}\Sigma^{\dagger}\partial_{\mu}\Sigma)
     -{\textstyle\frac{1}{4}}\mu^{2}{\rm Tr}(\Sigma^{\dagger}\Sigma)
     -{\textstyle\frac{1}{8}}\lambda^{2}{\rm Tr}
      (\Sigma^{\dagger}\Sigma)^{2},                          \nonumber\\
  {\cal L}_{I} &=& -g(\overline{\psi}_{L}\Sigma\psi_{R}
        +\overline{\psi}_{R}\Sigma^{\dagger}\psi_{L}),   \label{Lsu2lin}
\end{eqnarray}
and $\mu$, $\lambda$, and $g$ free parameters. Obviously, the ground
state cannot have the full SU(2)$_{L}$$\times$SU(2)$_{R}$ symmetry,
since in that case all hadrons (nucleon, pion, sigma in this model)
would have partners of equal mass but with opposite parity,
which is not the case in the real world. Hence, the chiral symmetry has
to be spontaneously broken down to the vectorial subgroup SU(2)$_{V}$.
This can be achieved by adding a term linear in the $\sigma$ field,
choosing the ground state as $\langle\sigma\rangle=f_{0}$, and shifting
the scalar field to $\sigma=s+f_{0}$. We then find the familiar
Lagrangian
\begin{eqnarray}
   {\cal L} &=& \overline{\psi}(i\gamma^{\mu}\partial_{\mu}-M)\psi
   -g\overline{\psi}(s+i\gamma_{5}\bbox{\tau}\!\cdot\!\bbox{\pi})\psi
                                                         \nonumber\\
   &&+{\textstyle\frac{1}{2}}(\partial^{\mu}s\partial_{\mu}s
       -m_{s}^{2}s^{2}) + {\textstyle\frac{1}{2}}
      (\partial^{\mu}\bbox{\pi}\!\cdot\!\partial_{\mu}\bbox{\pi}
       -m_{\pi}^{2}\bbox{\pi}^{2})                      \nonumber\\
   &&-\frac{m_{s}^{2}-m_{\pi}^{2}}{2f_{0}}\,s(s^{2}+\bbox{\pi}^{2})
     -\frac{m_{s}^{2}-m_{\pi}^{2}}{8f_{0}^{2}}\,
        (s^{2}+\bbox{\pi}^{2})^{2},                   \label{pscop}
\end{eqnarray}
where the nucleon mass is given by $M=gf_{0}$, and the scalar and
pseudoscalar masses are given by $m_{s}^{2}=\mu^{2}+3\lambda^{2}f_{0}^{2}$
and $m_{\pi}^{2}=\mu^{2}+\lambda^{2}f_{0}^{2}$, respectively.
The PCAC (partially-conserved axial-vector current) condition for the
axial-vector Noether current, $\partial_{\mu}{\bf A}^{\mu}=m_{\pi}^{2}
f_{\pi}\bbox{\pi}$, determines the relation between
$\langle\sigma\rangle=f_{0}$ and $f_{\pi}=92.4\pm0.3$ MeV~\cite{PDG96},
the pion decay constant.
As we will see below, the introduction of vector and axial-vector fields
enforces a renormalization of the pion field, and so we cannot make
the identification $f_{0}=f_{\pi}$ (a true identity in the absence of
vector and axial-vector gauge fields).

Weinberg has shown~\cite{Wei66} that the linear $\sigma$ model has
the major disadvantage that it hides the fact that soft pions are
emitted in clusters by derivative couplings from external lines,
and so it is more convenient to transform the nonderivative
$\overline{\psi}i\gamma_{5}\bbox{\tau}\!\cdot\!\bbox{\pi}\psi$ and
$\sigma\bbox{\pi}^{2}$ interactions away. For that purpose, Weinberg
defines a nonlinear transformation of the nucleon fields, given
by~\cite{Wei67}
\begin{equation}
     \psi = (1+\bbox{\xi}^{2})^{-1/2}
            (1-i\gamma_{5}\bbox{\tau}\!\cdot\!\bbox{\xi})N,
\end{equation}
with $\bbox{\xi}$ chosen such that
\begin{equation}
   \overline{\psi}(M+gs+ig\gamma_{5}\bbox{\tau}\!\cdot\!\bbox{\pi})\psi
   =\overline{N}(M+gs')N.                 \label{nogam5}
\end{equation}
It will be convenient to write this transformation in the form
\begin{equation}
    \left. \begin{array}{l}
    N_{R}=u\psi_{R} \\ N_{L}=u^{\dagger}\psi_{L}  \end{array}\
    \right\},\ \ u(\bbox{\xi})=\frac{1+i\bbox{\tau}\!\cdot\!\bbox{\xi}}
                     {(1+\bbox{\xi}^{2})^{1/2}},      \label{udef2}
\end{equation}
in which case we find
\begin{equation}
   \Sigma' \equiv f_{0}+s'=u^{\dagger}\Sigma u^{\dagger}.
\end{equation}
Solving Eq.~(\ref{nogam5}) gives
\begin{eqnarray}
   \bbox{\xi}&=&\bbox{\pi}\left[f_{0}+s
             +\sqrt{(f_{0}+s)^{2}+\bbox{\pi}^{2}}\right]^{-1}
              \equiv \frac{1}{2f_{0}}\bbox{\pi}',     \nonumber\\
   s'&=&[(f_{0}+s)^{2}+\bbox{\pi}^{2}]^{1/2}-f_{0}
        =(\Sigma^{\dagger}\Sigma)^{1/2}-f_{0},      \label{newspi}
\end{eqnarray}
where the chiral rotation vector $\bbox{\xi}$ is proportional to a new
pion field, $\bbox{\pi}'$. We should mention that the condition
(\ref{nogam5}) also removes the $\sigma\bbox{\pi}^{2}$ interaction,
and the new Lagrangian ${\cal L}_{\Sigma'}$ only contains three-point
and four-point interactions for the $s'$ scalar field.

In this nonlinear realization of the spontaneously broken chiral
symmetry, the new fields $u(\bbox{\xi})$ of the coset space
SU(2)$_{L}$$\times$SU(2)$_{R}$/SU(2)$_{V}$ transform as~\cite{Col69}
\begin{mathletters}
\begin{eqnarray}
   u &\rightarrow& LuH^{\dagger}=HuR^{\dagger},   \label{utrans}\\
   H &=& \sqrt{L u^{2} R^{\dagger}}R u^{\dagger}
        =\sqrt{R u^{\dagger2} L^{\dagger}}L u,      \label{Hdef}
\end{eqnarray}
\end{mathletters}
where the equality in Eq.~(\ref{utrans}) is due to parity.
The left and right components of the new doublet nucleon field $N$
both transform with the same SU(2) matrix $H$, and so the nucleon
part of the Lagrangian can now be written as
\begin{equation}
  {\cal L}_{N} = \overline{N}(i\gamma^{\mu}D_{\mu}-M)N - g\overline{N}Ns'
    -g_{A}\overline{N}\gamma^{5}\gamma^{\mu}u_{\mu}N, \label{Lsu2nonlin}
\end{equation}
where we defined $D_{\mu}=\partial_{\mu}+i\Gamma_{\mu}$, and
\begin{eqnarray}
   \Gamma_{\mu}&=&-{\textstyle\frac{i}{2}}\left(u^{\dagger}\partial_{\mu}u
                  +u\partial_{\mu}u^{\dagger}\right)
       =  \frac{1}{4f_{0}^{2}}\,
      \frac{\bbox{\tau}\!\cdot\!\bbox{\pi}'\times\partial_{\mu}\bbox{\pi}'}
           {1+\bbox{\xi}^{2}}, \nonumber\\
   u_{\mu}&=&-{\textstyle\frac{i}{2}}\left(u^{\dagger}\partial_{\mu}u
                  -u\partial_{\mu}u^{\dagger}\right)
       =  \frac{1}{2f_{0}}\,
       \frac{\bbox{\tau}\!\cdot\!\partial_{\mu}\bbox{\pi}'}
            {1+\bbox{\xi}^{2}}.
\end{eqnarray}
The transformation rule for the connection $\Gamma_{\mu}$,
\begin{equation}
    \Gamma_{\mu} \rightarrow H\Gamma_{\mu}H^{\dagger}
                     -iH\partial_{\mu}H^{\dagger},    \label{Gamtrans}
\end{equation}
ensures the invariance of the kinetic-energy term.
The transformation rule for the $N$ field, $N\rightarrow H\!N$, means
that $\overline{N}N$ is already an invariant in itself, and so the
mass $M$ can be treated as a free parameter.
Similarly, the transformation rule for $u_{\mu}$,
\begin{equation}
    u_{\mu} \rightarrow Hu_{\mu}H^{\dagger},        \label{umutrans}
\end{equation}
allows for the introduction of the free parameter $g_{A}$ in
Eq.~(\ref{Lsu2nonlin}).
Choosing it to be $g_{A}=1.2601\pm0.0025~\cite{PDG96}$, which is the
value for the weak interaction axial-vector coupling constant, and
substituting for $\bbox{\pi}'$, the pion is seen to couple to the
nucleon via the pseudovector coupling with strength $g_{A}/2f_{\pi}$.
This relation between the weak axial-vector and the pion pseudovector
coupling constants is known as the Goldberger-Treiman
relation~\cite{Gol58}. Although this relation only exactly holds in
the chiral limit, experimentally it is found to hold within about 2\%.
Note that in the $\psi$ representation of Eq.~(\ref{pscop}), this
coupling is found to be $g_{A}=1$, which is off by 25\%.

The next step is to add the isotriplets of vector ($\bbox{\rho}$) and
axial-vector (${\bf a}_{1}$) fields. One way to do this is to extend
the global chiral symmetry to a local one and to define the
corresponding left- and right-handed gauge fields as
\begin{eqnarray}
  l_{\mu}&\equiv&{\textstyle\frac{1}{2}}\bbox{\tau}\!\cdot\!{\bf l}_{\mu}
       ={\textstyle\frac{1}{2}}\bbox{\tau}\!\cdot\!
                     (\bbox{\rho}_{\mu}+{\bf a}_{\mu}),   \nonumber\\
  r_{\mu}&\equiv&{\textstyle\frac{1}{2}}\bbox{\tau}\!\cdot\!{\bf r}_{\mu}
       ={\textstyle\frac{1}{2}}\bbox{\tau}\!\cdot\!
                     (\bbox{\rho}_{\mu}-{\bf a}_{\mu}),
\end{eqnarray}
with their field strength tensors
\begin{equation}
   l_{\mu\nu}=\partial_{\mu}l_{\nu}-\partial_{\nu}l_{\mu}
              +ig_{V}[l_{\mu},l_{\nu}],     \label{fieldstrength}
\end{equation}
and similarly for $r_{\mu\nu}$. Their transformation properties are
\begin{eqnarray}
   l_{\mu} &\rightarrow& Ll_{\mu}L^{\dagger}
             -\frac{i}{g_{V}}L\partial_{\mu}L^{\dagger}, \ \ \ \ \
           l_{\mu\nu}\rightarrow Ll_{\mu\nu}L^{\dagger}, \nonumber\\
   r_{\mu} &\rightarrow& Rr_{\mu}R^{\dagger}
             -\frac{i}{g_{V}}R\partial_{\mu}R^{\dagger}, \ \ \
           r_{\mu\nu}\rightarrow Rr_{\mu\nu}R^{\dagger},
\end{eqnarray}
and so the proper field strength tensors for the nonlinear
transformation are $u^{\dagger}l_{\mu\nu}u$ and
$ur_{\mu\nu}u^{\dagger}$. Invariance under local chiral
transformations implies the extensions
\begin{eqnarray}
   \Gamma_{\mu}&=&-{\textstyle\frac{i}{2}}\left[u^{\dagger}
              (\partial_{\mu}+ig_{V}l_{\mu})u
        +u(\partial_{\mu}+ig_{V}r_{\mu})u^{\dagger}\right], \nonumber\\
   u_{\mu}&=&-{\textstyle\frac{i}{2}}\left[u^{\dagger}
              (\partial_{\mu}+ig_{V}l_{\mu})u
        -u(\partial_{\mu}+ig_{V}r_{\mu})u^{\dagger}\right].
\end{eqnarray}
The locally chiral-invariant Lagrangian can now be written
as~\cite{Wes67}
\begin{equation}
  {\cal L}_{N} = \overline{N}(i\gamma^{\mu}\partial_{\mu}-M)N
     -g_{s}\overline{N}Ns'-{\textstyle\frac{1}{2}}g_{V}
        \overline{N}\gamma^{\mu}\bbox{\tau}N\!\cdot\!\bbox{\rho}'_{\mu}
     -{\textstyle\frac{1}{2}}\lambda\overline{N}\gamma^{5}\gamma^{\mu}
        \bbox{\tau}N\!\cdot\!{\bf a}'_{\mu},     \label{Lsu2nonlinloc}
\end{equation}
where $g_{s}$, $g_{V}$, and $\lambda$ are free parameters, and where
we defined the field abbreviations
\begin{eqnarray}
  \bbox{\rho}'_{\mu} &=& \bbox{\rho}_{\mu}+\frac{2}{1+\bbox{\xi}^{2}}\,
     \bbox{\xi}\!\times\!{\bf a}_{\mu}+\frac{2}{g_{V}}
     \frac{1}{1+\bbox{\xi}^{2}}\,\bbox{\xi}\!\times\!(\partial_{\mu}
     \bbox{\xi}+g_{V}\bbox{\xi}\!\times\!\bbox{\rho}_{\mu}), \nonumber\\
  {\bf a}'_{\mu} &=& {\bf a}_{\mu}-\frac{2}{1+\bbox{\xi}^{2}}\,
     \bbox{\xi}\!\times\!({\bf a}_{\mu}\!\times\!\bbox{\xi})
     +\frac{2}{g_{V}}\frac{1}{1+\bbox{\xi}^{2}}\,(\partial_{\mu}
     \bbox{\xi}+g_{V}\bbox{\xi}\!\times\!\bbox{\rho}_{\mu}).
\end{eqnarray}
These field combinations are seen to give rise to pair ($\pi\pi$,
$\pi\rho$, $\pi a_{1}$) vertices, and other higher-order
multiple-meson vertices.

As an example of how this model can be made to agree with the actual
experimental results, we next give some attention to the meson sector.
The axial-vector field can be given a mass by adding to the Lagrangian
a chiral-invariant term proportional to ${\rm Tr}(u_{\mu}u^{\mu})$;
this will also generate the kinetic-energy term for the pion field.
The vector field receives its mass from a term proportional to
${\rm Tr}(l_{\mu}l^{\mu}+r_{\mu}r^{\mu})$, which is only invariant
under {\it global\/} chiral transformations. This term also contributes
to the mass of the axial-vector fields. However, in order to get an
acceptable agreement with experiment (not only with respect to the
empirical masses, but also with respect to some of the empirical meson
coupling constants and decay widths), we have to extend the Lagrangian
even further. A very general form in the context of the linear $\sigma$
model is given by Ko and Rudaz~\cite{Ko94}. Translating their results
in terms of the fields of the nonlinear transformation (\ref{udef2}),
the masses for the vector and axial-vector mesons are obtained from
the Lagrangian
\begin{equation}
  {\cal L}_{m} = f_{0}^{2}{\rm Tr}(u^{\mu}u_{\mu})
      + {\textstyle\frac{1}{2}}\,m_{0}^{2}{\rm Tr}(l^{\mu}l_{\mu}
         +r^{\mu}r_{\mu}) - c\,g_{V}^{2}(f_{0}+s')^{2}{\rm Tr}
        (l^{\mu}u^{2}r_{\mu}u^{\dagger2}).    \label{vecmass}
\end{equation}
We find that there is a mixing between the $\bbox{\pi}$ and
${\bf a}_{\mu}$ fields, which can be removed by making a field
redefinition for the ${\bf a}_{\mu}$ field, ${\bf a}_{\mu}={\bf A}_{\mu}
-h(\partial_{\mu}\bbox{\pi}+g_{V}\bbox{\pi}\!\times\!\bbox{\rho})$, and
then choosing $h$ appropriately.
Also, the kinetic-energy term for the pion field is no longer of the
canonical form, and so we have to do a wave function renormalization:
$\bbox{\pi}=Z_{\pi}^{-1/2}\bbox{\pi}_{r}$. The final result reads
(for details, see Ref.~\cite{Ko94})
\begin{eqnarray}
   m_{\rho}^{2} &=& m_{0}^{2}-c\,g_{V}^{2}f_{0}^{2},    \nonumber\\
   m_{a_{1}}^{2} &=& m_{0}^{2}+(c+1)g_{V}^{2}f_{0}^{2}, \nonumber\\
   Z_{\pi} &=& 1-g_{V}^{2}f_{0}^{2}/m_{a_{1}}^{2},  \label{Zpisolve}
\end{eqnarray}
and the PCAC condition gives $f_{\pi}=Z_{\pi}^{1/2}f_{0}=92.4$ MeV.
The parameter $c$ is needed to get a simultaneous agreement for the
$\rho$ mass ($m_{\rho}=770$ MeV), the $a_{1}$ mass ($m_{a_{1}}=1.23$
GeV), and the $\rho N\!N$ coupling constant ($g_{V}=5.04$ from
$\rho^{0}\rightarrow e^{+}e^{-}$); all data from Ref.~\cite{PDG96}.
Substituting $f_{0}$ in the equation for $Z_{\pi}$ leads to the two
solutions $(Z_{\pi},c)=(0.173,-0.131)$ and $(0.827,1.25)$.
Similarly, the numerical value for the $\rho\pi\pi$ coupling constant
($g_{\rho\pi\pi}=6.05$ from $\rho\rightarrow\pi^{+}\pi^{-}$~\cite{PDG96})
can be enforced by adding yet another term to the Lagrangian,
introducing a new parameter $\kappa_{6}$ (see Ref.~\cite{Ko94}).
Having fixed both $c$ and $\kappa_{6}$, some of the other meson
properties such as the root-mean-square pion radius and $a_{1}$ decay
widths also come out very favorably~\cite{Ko94}.

Although a full discussion of the meson sector is outside the scope
of this paper, it is important to note that it is indeed possible
to construct a chiral-invariant Lagrangian (sometimes at the expense of
introducing new parameters and small chiral-symmetry violating terms)
which does a reasonably good job in describing a variety of experimental
results. Furthermore, the introduction of the axial-vector field as
a gauge boson enforces a renormalization of the pion field, and so we
explicitly need part of the Lagrangian for the meson sector to define
this renormalization constant.

\section{SU(3) chiral symmetry}
\label{sec:su3}
The extension to SU(3) is easily obtained by replacing the Pauli
isospin matrices, $\tau_{i}$, by the Gell-Mann matrices, $\lambda_{a}$.
The two-component nucleon field is replaced by a 3$\times$3 traceless
baryon-octet matrix $\Psi$, where the left- and right-handed components
transform as
\begin{equation}
   \Psi_{L}\rightarrow L\Psi_{L}L^{\dagger}, \ \ \ \
   \Psi_{R}\rightarrow R\Psi_{R}R^{\dagger}.   \label{BLRtrans}
\end{equation}
The meson content of the model still consists of the isosinglet scalar,
$\sigma$, but the pseudoscalar triplet is extended to an octet by
including the $\eta_{8}$ and the four $K$ mesons. They are collectively
denoted by $\pi_{a}$. These nine mesons are grouped into
\begin{equation}
  \Sigma = \sigma + i\lambda_{a}\pi_{a}, \ \ \ (a=1,\ldots,8),
\end{equation}
which still transforms according to Eq.~(\ref{Sigmatrans}), but
where $L$ and $R$ are now elements of SU(3).

Translating the definition (\ref{udef2}) to the SU(3) case, we write
\begin{equation}
  \left. \begin{array}{l}
     B_{R}=u\Psi_{R}u^{\dagger} \\ B_{L}=u^{\dagger}\Psi_{L}u
         \end{array}\ \right\},\ \
  u(\xi_{a})=\exp[i\lambda_{a}\xi_{a}] \equiv
    \exp\left[\frac{i\lambda_{a}\pi'_{a}}{2f_{0}}\right], \label{udef3}
\end{equation}
where we used a more convenient representation for $u(\xi_{a})$, the
elements of the coset space SU(3)$_{L}$$\times$SU(3)$_{R}$/SU(3)$_{V}$.
We will identify the primed fields with the octet of physical
pseudoscalar fields, which is given by
\begin{equation}
  \frac{1}{\sqrt{2}}\lambda_{a}\pi'_{a} = \left( \begin{array}{ccc}
      {\displaystyle\frac{\pi^{0}}{\sqrt{2}}+\frac{\eta_{8}}{\sqrt{6}}}
             & \pi^{+}  &  K^{+}  \\[2mm]
      \pi^{-} & {\displaystyle-\frac{\pi^{0}}{\sqrt{2}}
         +\frac{\eta_{8}}{\sqrt{6}}}  &   K^{0} \\[2mm]
      K^{-}  &  \overline{{K}^{0}}
             &  {\displaystyle-\frac{2\eta_{8}}{\sqrt{6}}}
             \end{array} \right),    \label{PSmat}
\end{equation}
and where the fields transform according to the SU(3) analogue of
Eq.~(\ref{utrans}). Following the phase convention of Ref.~\cite{Pai66},
the new octet of baryon fields reads
\begin{equation}
  B = \left( \begin{array}{ccc}
      {\displaystyle\frac{\Sigma^{0}}{\sqrt{2}}+\frac{\Lambda}{\sqrt{6}}}
               &  \Sigma^{+}  &  p  \\[2mm]
      \Sigma^{-}  & {\displaystyle-\frac{\Sigma^{0}}{\sqrt{2}}
                     +\frac{\Lambda}{\sqrt{6}}}  &  n \\[2mm]
      \Xi^{-}  &  -\Xi^{0}
               &  {\displaystyle-\frac{2\Lambda}{\sqrt{6}}}
             \end{array} \right),
\end{equation}
which transforms as
\begin{equation}
   B \rightarrow  HBH^{\dagger}.   \label{Btrans}
\end{equation}
The slightly different transformation property of the baryon octet
matrix requires that the SU(3) analogue to the nucleon derivative,
$D_{\mu}N=(\partial_{\mu}+i\Gamma_{\mu})N$ in Eq.~(\ref{Lsu2nonlin}),
reads
\begin{equation}
    D_{\mu}B = \partial_{\mu}B + i[\Gamma_{\mu},B].   \label{Dcovar}
\end{equation}

Due to the transformation properties (\ref{BLRtrans}) of the original
octet fields, we cannot simply copy the interaction Lagrangian of
Eq.~(\ref{Lsu2lin}) to the SU(3) case. To ensure invariance under
chiral SU(3)$_{L}$$\times$SU(3)$_{R}$, the $\Sigma$ field would then
have to appear quadratically; i.e., the interaction Lagrangian has to be
of the form ${\rm Tr}(\overline{\Psi}_{L}\Sigma \Psi_{R}\Sigma^{\dagger}
+\overline{\Psi}_{R}\Sigma^{\dagger}\Psi_{L}\Sigma)$.
Alternatively, we note that the field combinations $u\Sigma^{\dagger}u$
and $u^{\dagger}\Sigma u^{\dagger}$ both transform according to
Eq.~(\ref{Btrans}). Defining the combination
\begin{equation}
  \chi_{\pm}={\textstyle\frac{1}{2}}
       \left(u^{\dagger}\Sigma u^{\dagger} \pm
             u\Sigma^{\dagger}u\right),           \label{chipm}
\end{equation}
we thus find that the most simple interaction Lagrangian is given by
\begin{equation}
    {\cal L}_{I}=-g_{s,1}\,{\rm Tr}(\overline{B}\chi_{+}B)
                 -g_{s,2}\,{\rm Tr}(\overline{B}B\chi_{+}), \label{Lsps}
\end{equation}
where $g_{s,1}$ and $g_{s,2}$ are arbitrary constants.
In principle, we also could have included a term of the form
$g_{p}{\rm Tr}(\overline{B}\gamma_{5}\chi_{-}B)$, but that would have
re-introduced a nonderivative pseudoscalar interaction~\footnote{This
in contrast to the SU(2) case, where we can choose $u(\xi_{i})$ such
that $\chi_{-}=0$, and so all nonderivative pseudoscalar interactions
are transformed away. The difference is due to the fact that
$(\tau_{i}\pi_{i})(\tau_{j}\pi_{j})$ is proportional to $\openone_{2}$,
whereas $(\lambda_{a}\pi_{a})(\lambda_{b}\pi_{b})$ is not proportional
to $\openone_{3}$.}, which we wanted to get rid of in the first place.

Unfortunately, the interaction Lagrangian (\ref{Lsps}) is not good
enough if we want it to generate the empirical baryon masses, since
it gives the same mass, $M=(g_{s,1}+g_{s,2})f_{0}$, for all baryons in
the baryon octet. This problem can be solved by adding an octet of
scalar fields, $\lambda_{a}\sigma_{a}$, where the isoscalar octet
member is given a nonvanishing vacuum expectation value. We can then
extend $\Sigma$ by writing ($\lambda_{0}=\sqrt{2/3}\openone_{3}$)
\begin{equation}
   \Sigma=F+\lambda_{0}s_{0}+\lambda_{a}s_{a}
       +i\lambda_{a}\pi_{a}, \ \ \ \ (a=1,\ldots,8), \label{chistart}
\end{equation}
with $\pi_{a}$ the {\it original\/} pseudoscalar fields and
$F$ the vacuum expectation value of the scalar-nonet fields, i.e.
\begin{equation}
     F=\left(  \begin{array}{ccc}
             f_{1} & 0 & 0 \\ 0 & f_{1} & 0 \\ 0 & 0 & f_{2}
               \end{array} \right),
\end{equation}
with
\begin{equation}
   f_{1} = {\textstyle\sqrt{\frac{2}{3}}}\langle\sigma_{0}\rangle
          +{\textstyle\sqrt{\frac{1}{3}}}\langle\sigma_{8}\rangle, \ \ \
   f_{2} = {\textstyle\sqrt{\frac{2}{3}}}\langle\sigma_{0}\rangle
         -2{\textstyle\sqrt{\frac{1}{3}}}\langle\sigma_{8}\rangle.
                                                 \label{VEV}
\end{equation}
In the Appendix we show that now it is still possible to find a
transformation $u(\xi)$, which transforms away the octet of original
pseudoscalar fields while leaving the vacuum expectation matrix $F$
invariant, and where the $\xi_{a}$ fields can be identified with an
octet of new pseudoscalar fields. As before, this transformation
generates the combinations $\chi_{\pm}$ of Eq.~(\ref{chipm}). Both
combinations contain the {\it original\/} pseudoscalar fields in a
complicated way. However, $\chi_{+}$ behaves like a set of scalar
fields, and so we can simply {\it define\/} these new fields to be the
physical scalar fields and drop any reference to the original scalar
fields. Clearly, to zeroth order in the new pseudoscalar fields, the
old and new scalar fields are the same. Dropping primes, the nonet of
new scalar fields is given by
\begin{equation}
    \chi_{+}=F+\lambda_{0}s_{0}+\lambda_{a}s_{a},
\end{equation}
where $s_{0}$ now denotes the new scalar singlet and the octet is
given by
\begin{equation}
  \frac{1}{\sqrt{2}}\lambda_{a}s_{a} = \left( \begin{array}{ccc}
      {\displaystyle\frac{a_{0}^{0}}{\sqrt{2}}+\frac{s_{8}}{\sqrt{6}}}
         &  a_{0}^{+}  &  \kappa^{+}  \\[2mm]
      a_{0}^{-} & {\displaystyle-\frac{a_{0}^{0}}{\sqrt{2}}
    +\frac{s_{8}}{\sqrt{6}}} & \kappa^{0} \\[2mm]
      \kappa^{-}  &  \overline{{\kappa}^{0}}
      &  {\displaystyle-\frac{2s_{8}}{\sqrt{6}}}
             \end{array} \right).   \label{Smat}
\end{equation}
The pure octet isoscalar $s_{8}$ and pure singlet $s_{0}$ are
mixed to give the physical $f_{0}(980)$ and $\varepsilon(760)$.
Similarly, $-i\chi_{-}$ can be identified as a new isosinglet
pseudoscalar field, not present before. Because it transforms
as $H(-i\chi_{-})H^{\dagger}$, it can be formally added to the
octet pseudoscalar matrix, which completes the nonet. It must be
remembered, however, that group transformations such as given for
instance in Eq.~(\ref{utrans}) and which involve different
transformation matrices, are only valid for the (traceless) octet
matrices.

As discussed for the SU(2) case at the end of the previous section,
also in the SU(3) case the global chiral symmetry can be easily extended
to a local chiral symmetry. The required gauge fields are now given
by two octets of combinations of vector and axial-vector fields.
The vector octet is given by
\begin{equation}
  \frac{1}{\sqrt{2}}\lambda_{a}\rho_{a} = \left( \begin{array}{ccc}
      {\displaystyle\frac{\rho^{0}}{\sqrt{2}}+\frac{\omega_{8}}{\sqrt{6}}}
             & \rho^{+}  &  K^{\ast+}  \\[2mm]
      \rho^{-} & {\displaystyle-\frac{\rho^{0}}{\sqrt{2}}
         +\frac{\omega_{8}}{\sqrt{6}}}  &   K^{\ast0} \\[2mm]
      K^{\ast-}  &  \overline{{K}^{\ast0}}
             &  {\displaystyle-\frac{2\omega_{8}}{\sqrt{6}}}
             \end{array} \right),    \label{Vmat}
\end{equation}
and a similar matrix for the axial-vector octet.
Adding a mass term of the form (\ref{vecmass}) breaks the local symmetry
and introduces a mixing between the pseudoscalar and axial-vector fields,
which requires a redefinition of the axial-vector fields. This in turn
means that the kinetic-energy term for the pseudoscalar fields is no
longer of the canonical form, which is taken care of by renormalizing the
pseudoscalar fields. Unfortunately, then it is not possible to get a
satisfactory agreement for both the vector and axial-vector masses
simultaneously. But also this problem can be solved. One way is to
introduce yet another term~\cite{Ko94} proportional to
${\rm Tr}(l_{\mu\nu}\Sigma r^{\mu\nu}\Sigma^{\dagger})$, which
renormalizes the vector and axial-vector fields. Another option is
to modify the term proportional to $m_{0}^{2}$ by inserting the
combinations $\Sigma\Sigma^{\dagger}$ and $\Sigma^{\dagger}\Sigma$.
The advantage of the latter extension is that it does not involve
a new parameter, while it does allow for a very satisfactory
description of both vector and axial-vector masses. Therefore, a
convenient spin-1 mass Lagrangian reads
\begin{equation}
  {\cal L}^{(1)}_{m} = f_{1}^{2}{\rm Tr}(u^{\mu}u_{\mu})
      + \frac{m_{0}^{2}}{2f_{1}^{2}}\,{\rm Tr}
           (l^{\mu}\Sigma\Sigma^{\dagger}l_{\mu}
           +r^{\mu}\Sigma^{\dagger}\Sigma r_{\mu}) - c\,g_{V}^{2}
        {\rm Tr}(l^{\mu}\Sigma r_{\mu}\Sigma^{\dagger}). \label{mass1}
\end{equation}
Using this Lagrangian, we find that the set $m_{0}=653$ MeV, $c=-0.131$,
$f_{1}=222$ MeV, and $f_{2}=290$ MeV gives very satisfactory results
for the $\rho$, $a_{1}$, $K^{\ast}$, and $K_{1}$ masses, as well as
for the pion and kaon decay constants. Given these parameters, the
renormalization constants are $Z_{\pi}=0.173$, $Z_{K}=0.225$, and
$Z_{\eta_{8}}=0.236$. Note that $(Z_{\pi},c)$ is the same as in the
SU(2) case, and so the good phenomenology in the $\rho$--$a_{1}$
sector~\cite{Ko94} is still valid.
To make a further connection with the real physical world, we next
also introduce isosinglet vector and axial-vector fields, which
complete the experimentally observed nonets. The isosinglet fields
are taken to be SU(3) invariants. This means that we can formally
include them on the diagonal in the respective octet matrix
representations.

Finally, to be complete, we should also list the kinetic-energy
and mass terms for the scalar and pseudoscalar fields.
The kinetic-energy term for the scalar fields is given by
${\rm Tr}\,(D^{\mu}\chi^{\dagger}_{+}D_{\mu}\chi_{+})$, with the
covariant derivative defined as in Eq.~(\ref{Dcovar}) to ensure the
chiral invariance. The kinetic-energy term for the pseudoscalar
fields is already contained in ${\rm Tr}(u^{\mu}u_{\mu})$.
The simplest possible mass term for the scalar fields is of the form
\begin{equation}
  {\cal L}^{(0)}_{m,s}=f_{1}^{3}{\rm Tr}(\chi_{+}A_{s})
       -c_{2}f_{1}^{2}{\rm Tr}(\chi^{\dagger}_{+}\chi_{+})
       -c_{4}{\rm Tr}(\chi^{\dagger}_{+}\chi_{+})^{2},  \label{mass0sc}
\end{equation}
where the first term with the diagonal matrix $A_{s}={\rm diag}(x,x,y)$
breaks the chiral symmetry. This term has to be included in order to
remove the terms linear in the $s_{0}$ and $s_{8}$ fields as
generated by the last two terms. This condition fixes $(x,y)$ in
terms of $(c_{2},c_{4})$. Fitting to the scalar masses as discussed
in Sec.~\ref{sec:intro}, we find $(c_{2},c_{4})=(7.62,-0.46)$.

To generate the octet pseudoscalar masses, we again need a term
that breaks the chiral symmetry. The singlet pseudoscalar mass
is generated by ${\rm Tr}(\chi^{\dagger}_{-}\chi_{-})$, which
is chiral invariant. Hence, an appropriate Lagrangian for the
pseudoscalar meson masses looks like
\begin{equation}
  {\cal L}^{(0)}_{m,p}={\textstyle\frac{1}{4}}f_{1}^{2}
       {\rm Tr}\left[(u^{2}+u^{\dagger2})A_{p}\right]
     - {\textstyle\frac{1}{4}}m^{2}_{\eta_{0}}
       {\rm Tr}(\chi^{\dagger}_{-}\chi_{-}),    \label{mass0ps}
\end{equation}
with the diagonal matrix $A_{p}={\rm diag}(x',x',y')$, where
$x'=m^{2}_{\pi}/Z_{\pi}$ and $y'=2m^{2}_{K}/Z_{K}-m^{2}_{\pi}/Z_{\pi}$.
Note that with this choice the quadratic Gell-Mann--Okubo mass formula,
$3m^{2}_{\eta_{8}}+m^{2}_{\pi}-4m^{2}_{K}=0$, is approximately
satisfied. The mass $m_{\eta_{0}}$ is chosen such that, with the
proper mixing angle, we get the empirical $\eta$ and $\eta'$ masses.

The main achievement of the field transformations as discussed above
is that we have constructed various 3$\times$3 matrices of the general
form $\Phi=(1/\sqrt{2})\lambda_{c}\phi_{c}$, where $c=0,\ldots,8$.
These matrices contain scalar, pseudoscalar, vector, and axial-vector
meson fields, and (except for the vector mesons) they all transform
in the same way as the baryon fields. This allows us to define the
following chiral-invariant combinations
\begin{eqnarray}
  \left[\overline{B}B\Phi\right]_{F} &=& {\rm Tr}(\overline{B}\Phi B)
           -{\rm Tr}(\overline{B}B\Phi),  \nonumber\\
  \left[\overline{B}B\Phi\right]_{D} &=& {\rm Tr}(\overline{B}\Phi B)
           +{\rm Tr}(\overline{B}B\Phi)-{\textstyle\frac{2}{3}}\,
            {\rm Tr}(\overline{B}B){\rm Tr}(\Phi), \nonumber\\
  \left[\overline{B}B\Phi\right]_{S} &=&
            {\rm Tr}(\overline{B}B){\rm Tr}(\Phi).
\end{eqnarray}
For a traceless matrix $\Phi$ (i.e., only octets of mesons), the
$S$-type coupling $[\overline{B}B\Phi]_{S}$ vanishes, and the $F$- and
$D$-type couplings can be written as ${\rm Tr}(\overline{B}[\Phi,B])$
and ${\rm Tr} (\overline{B}\{\Phi,B\})$, respectively; a notation
often encountered in the literature.
Defining a baryon-baryon-meson octet coupling constant $g^{\rm oct}$
and a baryon-baryon-singlet coupling constant $g^{\rm sin}$, a
general interaction Lagrangian which satisfies chiral symmetry
can now be written in the form
\begin{equation}
   {\cal L}_{I} = -g^{\rm oct}\sqrt{2}\left\{
     \alpha\left[\overline{B}B\Phi\right]_{F}+
     (1-\alpha)\left[\overline{B}B\Phi\right]_{D}\right\}\, - \,
     g^{\rm sin}{\textstyle\sqrt{\frac{1}{3}}}
     \left[\overline{B}B\Phi\right]_{S},             \label{LIsu3}
\end{equation}
where $\alpha$ is known as the $F/(F+D)$ ratio, and the square-root
factors are introduced for later convenience.
The various field matrices are given by
\begin{mathletters}
\begin{eqnarray}
  \Phi_{\rm sc} &=& \frac{1}{\sqrt{2}}\left[
     F+\lambda_{c}s_{c}\right],                        \label{phisc}\\
  \Phi_{\rm vc} &=& \frac{-i}{\sqrt{2}g_{V}}\gamma_{\mu}\left[
     u^{\dagger}\left(\partial^{\mu}+{\textstyle\frac{i}{2}}g_{V}
     \lambda_{c}(\rho^{\mu}+A^{\mu}-hD^{\mu}\pi)_{c}\right)u\right.
            \nonumber\\ &&\hspace{50pt} \left.
     +u\left(\partial^{\mu}+{\textstyle\frac{i}{2}}g_{V}\lambda_{c}
     (\rho^{\mu}-A^{\mu}+hD^{\mu}\pi)_{c}\right)u^{\dagger}\right],
                                                       \label{phivc}\\
  \Phi_{\rm ax} &=& \frac{-i}{\sqrt{2}g_{V}}\gamma_{5}\gamma_{\mu}\left[
     u^{\dagger}\left(\partial^{\mu}+{\textstyle\frac{i}{2}}g_{V}
     \lambda_{c}(\rho^{\mu}+A^{\mu}-hD^{\mu}\pi)_{c}\right)u\right.
            \nonumber\\ &&\hspace{60pt} \left.
     -u\left(\partial^{\mu}+{\textstyle\frac{i}{2}}g_{V}\lambda_{c}
     (\rho^{\mu}-A^{\mu}+hD^{\mu}\pi)_{c}\right)u^{\dagger}\right],
                                                       \label{phiax}
\end{eqnarray}
\end{mathletters}
where $D^{\mu}(\lambda\pi)=\partial^{\mu}(\lambda\pi)-
{\textstyle\frac{i}{2}}g_{V}[(\lambda\pi),(\lambda\rho_{\mu})]$,
and $h$ chosen such that the mixing between the axial-vector and
pseudoscalar fields in the meson sector vanishes.
Note that the pseudovector coupling of the pseudoscalar fields is
already included in the axial-vector field matrix $\Phi_{\rm ax}$.

In addition to the electric coupling $\gamma^{\mu}\rho'_{\mu}$
[where $\rho'_{\mu}$ is a shorthand for the fields appearing in
Eq.~(\ref{phivc})], it is also possible~\cite{Wes67} to include a
chiral-invariant magnetic coupling $\sigma^{\mu\nu}\rho'_{\mu\nu}$,
where $\rho'_{\mu\nu}$ is the field strength tensor for the
$\rho'_{\mu}$ field combination. This is due to the fact that
$\rho'_{\mu}$ transforms according to Eq.~(\ref{Gamtrans}), and so we
can define a chiral-invariant field strength tensor $\rho'_{\mu\nu}$,
as in Eq.~(\ref{fieldstrength}).
The transformation (\ref{Gamtrans}) also imposes the constraint that
the chiral-variant $D$-type coupling $[\overline{B}B\Phi_{\rm vc}]_{D}$
should vanish, i.e., the electric $\alpha^{e}_{V}=1$. This represents
the so-called universality condition proposed by Sakurai~\cite{Sak60}.
Hence, the assumption that the $\rho$ meson couples universally to
the isospin current in this model is a direct consequence of chiral
SU(3) symmetry. On the other hand, the magnetic $\alpha^{m}_{V}$ is
still a free parameter.

Until now, it appears that we have not gained much by imposing
the SU(3)$\times$SU(3) chiral symmetry: the form of the interaction
Lagrangian (\ref{LIsu3}) can be written down immediately by assuming
only an SU(3) symmetry, and has been known for a long time (see, e.g.,
Ref.~\cite{Swa63}). However, the important difference is that the
3$\times$3 matrices $\Phi$ are not just simple representations of
the scalar, vector, or axial-vector meson fields, but they contain
the pseudoscalar fields in a nonlinear way as well. This means that
the chiral Lagrangian contains all kinds of multiple-meson (pair,
triple, etc.) interactions not envisaged before. The coupling
constants for these multiple-meson interactions can all be expressed
in terms of the single-meson interaction coupling constants. This
will be the subject of Sec.~\ref{sec:double-meson}.
Furthermore, the pseudoscalar fields are renormalized due to the
introduction of the axial-vector fields. Hence, already in leading
order the chiral Lagrangian gives interactions between baryons and
pseudoscalar mesons that are slightly different from what is obtained
with the standard (i.e., nonchiral) Lagrangian. Also, the coupling
constants of the pseudoscalar mesons are directly related to those
of the axial-vector mesons. Finally, the imposed symmetry allows us
to fit all the octet meson masses with only two parameters for each
octet.

\section{Single-meson vertices}
\label{sec:single-meson}
We will first look at the single-meson baryon-baryon coupling
constants, i.e., the interaction terms linear in the meson fields.
Let us drop for a moment the Lorentz character of the interaction
vertices ($\openone_{4}$ in spinor space for scalar mesons,
$\gamma_{5}\gamma_{\mu}\partial^{\mu}$ for pseudoscalar mesons,
$\gamma_{\mu}$ and $\sigma_{\mu\nu}\partial^{\nu}$ for vector mesons,
and $\gamma_{5}\gamma_{\mu}$ for axial-vector mesons) and take as
an example the nonet of pseudoscalar mesons.
The derivative (pseudovector-coupled) pseudoscalar-meson interaction
Lagrangian to lowest order in the fields is then of the form
\begin{equation}
   {\cal L}_{\rm pv}={\cal L}_{\rm pv}^{\{1\}}
                    +{\cal L}_{\rm pv}^{\{8\}},
\end{equation}
where the $S$-type coupling in Eq.~(\ref{LIsu3}) gives the singlet
interaction Lagrangian
\begin{equation}
   m_{\pi}{\cal L}_{\rm pv}^{\{1\}}=
       -f_{N\!N\eta_{0}}(\overline{N}N)\eta_{0}
       -f_{\Xi\Xi\eta_{0}}(\overline{\Xi}\Xi)\eta_{0}
       -f_{\Sigma\Sigma\eta_{0}}(\overline{\bbox{\Sigma}}\!\cdot\!
        \bbox{\Sigma})\eta_{0}
       -f_{\Lambda\Lambda\eta_{0}}(\overline{\Lambda}\Lambda)\eta_{0},
                                   \label{Lbar1}
\end{equation}
with the (derivative) pseudovector coupling constants
\begin{equation}
   f_{N\!N\eta_{0}}=f_{\Xi\Xi \eta_{0}}=f_{\Sigma\Sigma \eta_{0}}=
   f_{\Lambda\Lambda \eta_{0}}=f^{\rm sin}_{\rm pv},  \label{gsin}
\end{equation}
and where we introduced the charged-pion mass as a scaling mass to
make the pseudovector coupling $f$ dimensionless. The interaction
Lagrangian for the meson octet is obtained by evaluating the $F$- and
$D$-type couplings in Eq.~(\ref{LIsu3}), and can be written as
\begin{eqnarray}
   m_{\pi}{\cal L}_{\rm pv}^{\{8\}} &=&
  -f_{N\!N\pi}(\overline{N}\bbox{\tau}N)\!\cdot\!\bbox{\pi}
  -f_{\Xi\Xi\pi}(\overline{\Xi}\bbox{\tau}\Xi)\!\cdot\!\bbox{\pi}
  -f_{\Lambda\Sigma\pi}(\overline{\Lambda}\bbox{\Sigma}+
      \overline{\bbox{\Sigma}}\Lambda)\!\cdot\!\bbox{\pi}
  +if_{\Sigma\Sigma\pi}(\overline{\bbox{\Sigma}}\!\times\!\bbox{\Sigma})
      \!\cdot\!\bbox{\pi}             \nonumber\\
 &&-f_{N\!N\eta_{8}}(\overline{N}N)\eta_{8}
   -f_{\Xi\Xi\eta_{8}}(\overline{\Xi}\Xi)\eta_{8}
   -f_{\Lambda\Lambda\eta_{8}}(\overline{\Lambda}\Lambda)\eta_{8}
   -f_{\Sigma\Sigma\eta_{8}}(\overline{\bbox{\Sigma}}\!\cdot\!
       \bbox{\Sigma})\eta_{8}            \nonumber\\
 &&-f_{\Lambda N\!K}\left[(\overline{N}K)\Lambda
         +\overline{\Lambda}(\overline{K}N)\right]
   -f_{\Xi\Lambda K}\left[(\overline{\Xi}K_{c})\Lambda
         +\overline{\Lambda}(\overline{K_{c}}\Xi)\right] \nonumber\\
 &&-f_{\Sigma N\!K}\left[\overline{\bbox{\Sigma}}\!\cdot\!
         (\overline{K}\bbox{\tau}N)+(\overline{N}\bbox{\tau}K)
         \!\cdot\!\bbox{\Sigma}\right]
   -f_{\Xi\Sigma K}\left[\overline{\bbox{\Sigma}}\!\cdot\!
       (\overline{K_{c}}\bbox{\tau}\Xi)
     +(\overline{\Xi}\bbox{\tau}K_{c})\!\cdot\!\bbox{\Sigma}\right].
                              \label{Lbar8}
\end{eqnarray}
Here we introduced the doublets
\begin{equation}
  N=\left(\begin{array}{c} p \\ n \end{array} \right), \ \ \
  \Xi=\left(\begin{array}{c} \Xi^{0} \\ \Xi^{-} \end{array} \right), \ \ \
  K=\left(\begin{array}{c} K^{+} \\ K^{0} \end{array} \right),
  \ \ \   K_{c}=\left(\begin{array}{c} \overline{K^{0}} \\
               -K^{-} \end{array} \right),        \label{doublets}
\end{equation}
and $\bbox{\Sigma}$ and $\bbox{\pi}$ are isovectors with phases
chosen~\cite{Swa63} such that
\begin{equation}
  \bbox{\Sigma}\!\cdot\!\bbox{\pi} = \Sigma^{+}\pi^{-}
       +\Sigma^{0}\pi^{0}+\Sigma^{-}\pi^{+}.
\end{equation}
The octet coupling constants are given by the following expressions
($f\equiv f^{\rm oct}_{\rm pv}$)
\begin{equation}
  \begin{array}{lll}
    f_{N\!N\pi}=f, \ \ \ &
    f_{N\!N\eta_{8}}=\frac{1}{\sqrt{3}}\,f(4\alpha-1), \ \ \ &
    f_{\Lambda N\!K}=-\frac{1}{\sqrt{3}}\,f(1+2\alpha), \\[2mm]
    f_{\Xi\Xi\pi}=-f(1-2\alpha), \ \ \ &
    f_{\Xi\Xi\eta_{8}}=-\frac{1}{\sqrt{3}}\,f(1+2\alpha), \ \ \ &
    f_{\Xi\Lambda K}=-\frac{1}{\sqrt{3}}\,f(4\alpha-1), \\[2mm]
    f_{\Lambda\Sigma\pi}=\frac{2}{\sqrt{3}}\,f(1-\alpha), \ \ \ &
    f_{\Lambda\Lambda\eta_{8}}=-\frac{2}{\sqrt{3}}\,f(1-\alpha), \ \ \ &
    f_{\Sigma N\!K}=f(1-2\alpha), \\[2mm]
    f_{\Sigma\Sigma\pi}=2f\alpha, \ \ \ &
    f_{\Sigma\Sigma\eta_{8}}=\frac{2}{\sqrt{3}}\,f(1-\alpha), \ \ \ &
    f_{\Xi\Sigma K}=f.
  \end{array}           \label{goct}
\end{equation}
Similar relations (without the scaling mass $m_{\pi}$) are found
for the coupling constants of the scalar, vector, and axial-vector
mesons. Note, however, that for the pseudoscalar mesons the relations
(\ref{gsin}) and (\ref{goct}) should be slightly modified due to
the difference in renormalization factors $Z_{\pi}$, $Z_{K}$,
$Z_{\eta}$, and $Z_{\eta'}$ (see Sec.~\ref{sec:su3}).

For each type of meson there are only four parameters: the
singlet coupling, the octet coupling, the $F/(F+D)$ ratio, and
the mixing angle to generate the physical isoscalar mesons from
the pure octet and singlet fields. In most cases we can impose
theoretical and experimental constraints on these parameters.
This will be discussed next.
As already mentioned before, the axial-vector mesons are very heavy,
and hence are not expected to play an important role in low-energy
potential models, but we will still discuss them here for reasons of
completeness.

\subsection{Scalar mesons}
\label{subsec:scalar}
Because the existence of a nonet of scalar mesons with masses below
1 GeV is still controversial, the constraints on the SU(3) parameters
solely depend on the particular theoretical model one wants to
use to describe the scalar mesons. As a matter of fact, it is not
at all clear whether it is indeed valid to impose an SU(3) symmetry.
However, in order to limit the number of free parameters, we will
here assume the standard SU(3) relations and assume that the physical
$\varepsilon(760)$ and $f_{0}(980)$ mesons are admixtures of the pure
octet $\sigma_{8}$ and pure singlet $\sigma_{0}$ fields, in terms of
the scalar mixing angle $\theta_{S}$,
\begin{eqnarray}
  |f_{0}\rangle       &=& \sin\theta_{S}\,|\sigma_{8}\rangle
             +\cos\theta_{S}\,|\sigma_{0}\rangle,    \nonumber\\
  |\varepsilon\rangle &=& \cos\theta_{S}\,|\sigma_{8}\rangle
             -\sin\theta_{S}\,|\sigma_{0}\rangle.   \label{mixs}
\end{eqnarray}
A possible value for the mixing angle is obtained by assuming that
the scalar mesons are $\bar{q}^{2}q^{2}$ states~\cite{Jaf77}, and
that the $\varepsilon(760)$ does not contain any strange quarks
(hence, its low mass). This implies an ideal mixing angle
$\theta_{S}=35.3^{\circ}$.

A value for $\alpha_{S}$ can be determined by assuming that the
baryon masses are generated by the nonvanishing vacuum expectation
value of the $\sigma_{0}$ and $\sigma_{8}$ scalar fields.
Defining the vacuum expectation values $f_{1}$ and $f_{2}$ as in
Eq.~(\ref{VEV}), we have
\begin{equation}
    \sqrt{3}\langle \sigma_{0}\rangle =
    {\textstyle\sqrt{\frac{1}{2}}}\,(2f_{1}+f_{2}), \ \ \
    \sqrt{3}\langle \sigma_{8}\rangle = (f_{1}-f_{2}),
\end{equation}
or, assuming the ideal mixing $\theta_{S}=35.3^{\circ}$,
\begin{equation}
    f_{1}=\langle f_{0}(980)\rangle, \ \ \
    f_{2}=-\sqrt{2}\langle \varepsilon(760)\rangle.
\end{equation}
{}From the interaction Lagrangians (\ref{Lbar1}) and (\ref{Lbar8})
for the scalar fields we find the following relations for the baryon
masses,
\begin{eqnarray}
   M_{N}       &=& M_{0}-{\textstyle\frac{1}{3}}g^{\rm oct}_{\rm sc}
                         (4\alpha_{S}-1)(f_{2}-f_{1}),  \nonumber\\
   M_{\Lambda} &=& M_{0}-{\textstyle\frac{2}{3}}g^{\rm oct}_{\rm sc}
                         (\alpha_{S}-1)(f_{2}-f_{1}),   \nonumber\\
   M_{\Sigma}  &=& M_{0}+{\textstyle\frac{2}{3}}g^{\rm oct}_{\rm sc}
                         (\alpha_{S}-1)(f_{2}-f_{1}),   \nonumber\\
   M_{\Xi}     &=& M_{0}+{\textstyle\frac{1}{3}}g^{\rm oct}_{\rm sc}
                         (2\alpha_{S}+1)(f_{2}-f_{1}),
\end{eqnarray}
with $M_{0}=g^{\rm sin}_{\rm sc}\,(2f_{1}+f_{2})/\sqrt{6}=
{\textstyle\frac{1}{2}}(M_{\Lambda}+M_{\Sigma})$. According to the
relations given above, these masses satisfy the equality
$2M_{N}+2M_{\Xi}=3M_{\Lambda}+M_{\Sigma}$, which is indeed
approximately true experimentally (4516 MeV versus 4537 MeV).
Solving for $\alpha_{S}$, we find $\alpha_{S}=1.42$ and
$g^{\rm oct}_{\rm sc}(f_{2}-f_{1})=136.5$ MeV.

Finally, using the estimates for $f_{1}$ and $f_{2}$ as given in
Sec.~\ref{sec:su3}, we find $g^{\rm sin}_{\rm sc}=3.8$ and
$g^{\rm oct}_{\rm sc}=2.0$.

\subsection{Pseudoscalar mesons}
\label{subsec:pseudoscalar}
The mixing angle $\theta_{PS}$ for the pseudoscalar mesons is
defined by
\begin{eqnarray}
  |\eta \rangle &=& \cos\theta_{PS}\,|\eta_{8}\rangle
                   -\sin\theta_{PS}\,|\eta_{0}\rangle,    \nonumber\\
  |\eta'\rangle &=& \sin\theta_{PS}\,|\eta_{8}\rangle
                   +\cos\theta_{PS}\,|\eta_{0}\rangle.  \label{mixps}
\end{eqnarray}
The linear and quadratic Gell-Mann--Okubo mass formulas give~\cite{PDG96}
$\theta_{PS}\approx-23^{\circ}$ and $\theta_{PS}\approx-10^{\circ}$,
respectively. The current experimental evidence, however, seems to
favor~\cite{Gil87} $\theta_{PS}\approx-20^{\circ}$.

The axial-vector current $F/D$ ratio, obtained from the Cabibbo theory
of semileptonic decays of baryons, gives~\cite{Clo93} 0.575$\pm$0.0165,
or $\alpha_{PS}=0.365\pm0.007$.

The $\pi N\!N$ pseudovector coupling constant is obtained from the
Goldberger-Treiman relation~\cite{Gol58}, which gives
$f^{\rm oct}_{\rm pv}=f_{N\!N\pi}=g_{A}(0)m_{\pi}/2f_{\pi}=
0.952\pm0.003$. It can also be extracted from multienergy partial-wave
analyses of $pp$, $np$, and $\bar{p}p$ scattering data~\cite{Sto93b},
which yields for the coupling constant at the pion pole
$f_{N\!N\pi}^{2}/4\pi=0.0745\pm0.0006$. This latter result has to be
extrapolated to $t=0$ before comparing it with the Goldberger-Treiman
result.

Finally, we can also give an estimate for the singlet pseudovector
coupling constant. The estimate is based on the fact that we can
rewrite the $\eta_{8}$ and $\eta_{0}$ in a nonstrange-strange basis,
rather than the standard $\{u,d,s\}$ quark basis, and then assume
that the purely strange-quark state does not couple to the nucleon.
To be specific, in the standard quark basis
\begin{eqnarray}
   |\eta_{8}\rangle &=& {\textstyle\sqrt{\frac{1}{6}}}
                        |\bar{u}u+\bar{d}d-2\bar{s}s\rangle, \nonumber\\
   |\eta_{0}\rangle &=& {\textstyle\sqrt{\frac{1}{3}}}
                        |\bar{u}u+\bar{d}d+\bar{s}s\rangle,
\end{eqnarray}
and so the purely strange-quark state can be expressed as $|S\rangle=
|\bar{s}s\rangle=(|\eta_{0}\rangle-\sqrt{2}|\eta_{8}\rangle)/\sqrt{3}$.
This state does not couple to the nucleons provided that
$f_{N\!N\eta_{0}}=\sqrt{2}\,f_{N\!N\eta_{8}}$ or, equivalently,
$f^{\rm sin}_{\rm pv}={\textstyle\sqrt{\frac{2}{3}}}(4\alpha_{PS}-1)
f^{\rm oct}_{\rm pv}$.

\subsection{Vector mesons}
\label{subsec:vector}
The mixing angle $\theta_{V}$ for the vector mesons is defined by
\begin{eqnarray}
  |\omega\rangle &=& \sin\theta_{V}\,|\omega_{8}\rangle
                    +\cos\theta_{V}\,|\omega_{0}\rangle,   \nonumber\\
  |\phi  \rangle &=& \cos\theta_{V}\,|\omega_{8}\rangle
                    -\sin\theta_{V}\,|\omega_{0}\rangle.  \label{mixv}
\end{eqnarray}
Ideal mixing (i.e., the $\phi$ meson is a pure $\bar{s}s$ state)
gives $\theta_{V}=35.3^{\circ}$, which is very close to the
experimental value ($\theta_{V}\approx35^{\circ}$) and the values
found for the linear ($\theta_{V}\approx36^{\circ}$) and quadratic
($\theta_{V}\approx39^{\circ}$) Gell-Mann--Okubo mass
formulas~\cite{PDG96}.

As mentioned earlier, the transformation property of the connection
$\Gamma_{\mu}$ requires that $\alpha_{V}^{e}=1$ for the electric
coupling of the vector mesons, which is the universality
assumption~\cite{Sak60}.

The $\rho N\!N$ coupling constant is obtained from the electromagnetic
decay $\rho^{0}\rightarrow e^{+}e^{-}$. The vector-meson dominance
(VMD) hypothesis assumes that this decay proceeds via the photon
and that the $\rho^{0}$--$\gamma$ coupling is proportional to the
$\rho N\!N$ coupling~\cite{Sak60}. The decay width~\cite{PDG96}
$\Gamma(\rho^{0}\rightarrow e^{+}e^{-})=6.77\pm0.32$ keV then gives
$g^{\rm oct}_{\rm vc}=g_{N\!N\rho}=2.52\pm0.06$. Note the factor of
2 difference between the definition of $g^{\rm oct}_{\rm vc}$ in
Eq.~(\ref{LIsu3}) and $g_{V}$ in Eq.~(\ref{phivc}).

If we assume that the $\phi$ meson is a pure $\bar{s}s$ state, which
does not couple to the nucleons (i.e., an ideal mixing angle
$\theta_{V}=35.3^{\circ}$), then we find for the singlet coupling
$g^{\rm sin}_{\rm vc}=\sqrt{6}g^{\rm oct}_{\rm vc}$ or, equivalently,
$g_{N\!N\omega}=3g_{N\!N\rho}$. Perhaps a better estimate is obtained
from the decay width~\cite{PDG96} $\Gamma(\omega\rightarrow e^{+}e^{-})
=0.60\pm0.02$ keV, which  suggests $g_{N\!N\omega}/g_{N\!N\rho}=
\left[(m_{\omega}\Gamma_{\rho^{0}\rightarrow e^{+}e^{-}})/(m_{\rho}
\Gamma_{\omega\rightarrow e^{+}e^{-}})\right]^{1/2}=3.4\pm0.1$.

Keeping only the leading-order contribution proportional to
$\sigma_{\mu\nu}\partial^{\nu}\rho^{\mu}$, the magnetic coupling of
the vector mesons is defined as $f^{\rm oct}_{\rm vc}/2{\cal M}$,
where the scaling mass ${\cal M}$, taken to be the proton mass,
is included to make $f^{\rm oct}_{\rm vc}$ dimensionless.
Following Ref.~\cite{Sak65}, the SU(6) result for $\alpha_{V}^{m}$ can
be expressed as $\alpha_{V}^{m}=(4M_{8}-m_{v8})/(10M_{8}+2m_{v8})$,
where $M_{8}$ denotes the average mass in the baryon octet and
$m_{v8}$ the average mass in the vector-meson octet. This gives
$\alpha_{V}^{m}\approx0.28$ for the relativistic SU(6) case,
while $\alpha_{V}^{m}={\textstyle\frac{2}{5}}$ for the static
SU(6) case.

Again applying the VMD hypothesis and assuming that the lowest-mass
vector mesons ($\rho$, $\omega$, $\phi$) saturate the nucleon
electromagnetic form factors, the magnetic couplings are
given in terms of the anomalous magnetic moments of the proton and
neutron. This gives $(f/g)_{N\!N\rho}=\kappa_{p}-\kappa_{n}=3.71$, and
the isoscalar values are expected to be close to $(f/g)_{N\!N\omega}
+(f/g)_{N\!N\phi}\approx\kappa_{p}+\kappa_{n}=-0.12$.

\subsection{Axial-vector mesons}
\label{subsec:axial-vector}
We follow Ref.~\cite{Coo96} in estimating the value for the mixing
angle $\theta_{A}$, defined by
\begin{eqnarray}
  |f_{1}(1285)\rangle &=& \cos\theta_{A}\,|a_{8}\rangle
                         -\sin\theta_{A}\,|a_{0}\rangle,    \nonumber\\
  |f_{1}(1420)\rangle &=& \sin\theta_{A}\,|a_{8}\rangle
                         +\cos\theta_{A}\,|a_{0}\rangle.  \label{mixax}
\end{eqnarray}
For that purpose we rewrite the $f_{1}(1285)$ and $f_{1}(1420)$ mesons
in the nonstrange-strange basis,
\begin{eqnarray}
  |f_{1}(1285)\rangle &=& \cos\phi_{A}\,|A_{NS}\rangle
                         -\sin\phi_{A}\,|A_{S}\rangle,    \nonumber\\
  |f_{1}(1420)\rangle &=& \sin\phi_{A}\,|A_{NS}\rangle
                         +\cos\phi_{A}\,|A_{S}\rangle.
\end{eqnarray}
The observed decay widths give a mixing angle $\phi_{A}\approx
12^{\circ}$. In the singlet-octet basis the mixing angle is then
given by $\theta_{A}=\phi_{A}-\arctan\sqrt{2}=-42.7^{\circ}$.

The axial-vector coupling constants are closely related to the
pseudovector coupling constants due to the mixing between the
axial-vector and pseudoscalar fields; see the end of Sec.~\ref{sec:su2}.
The redefinition of the axial-vector fields and the renormalization
of the pseudoscalar fields gives the relation between the
coupling constants as
\begin{equation}
   g^{\rm oct}_{\rm ax} = g^{\rm oct}_{\rm vc}\,\frac{g_{A}(0)}{Z_{\pi}}
   = g^{\rm oct}_{\rm vc} g_{A}(0)
     \left(1-\frac{g_{V}^{2}f_{1}^{2}}{m_{a_{1}}^{2}}\right)^{-1}.
\end{equation}
The above relation can be recast into the form
\begin{equation}
   g_{N\!Na_{1}}=\frac{m_{a_{1}}}{m_{\pi}}f_{N\!N\pi}\,
       \sqrt{\frac{1-Z_{\pi}}{Z_{\pi}}},  \label{ga1tofpi}
\end{equation}
where the square root equals 1 for $Z_{\pi}=1/2$. The latter choice
gives the familiar result~\cite{Wei67,Wes67,Sch67} $g_{N\!Na_{1}}=
(m_{a_{1}}/m_{\pi})f_{N\!N\pi}\approx8.4$. But this assumes $m_{a_{1}}
=\sqrt{2}m_{\rho}$, which is not supported by experiment.
With our previous choice $(Z_{\pi},c)=(0.173,-0.131)$, the
axial-vector coupling constant comes out to be much larger,
$g^{\rm oct}_{\rm ax}\approx18$.
A more moderate value is obtained by choosing the alternative
solution to Eq.~(\ref{Zpisolve}), $(Z_{\pi},c)=(0.827,1.25)$,
which gives $g^{\rm oct}_{\rm ax}\approx3.8$.
An estimate for the $a_{1}N\!N$ coupling constant from experiment is
based on the idea of axial-vector meson dominance, which relates it
to the axial-vector coupling constant of the weak interaction,
$g_{A}(0)$, and the decay constant $f_{a_{1}}$, as defined by the
isovector $a_{1}$-to-vacuum matrix element of the hadronic axial-vector
current. With our definition (\ref{LIsu3}) this gives~\cite{Coo96}
$g^{\rm oct}_{\rm ax}=4.7\pm0.6$. Hence, we find contradictory
results: the larger $|c|$ solution gives reasonable agreement
with the empirical $g_{N\!Na_{1}}$ coupling constant, whereas the
smaller $|c|$ solution gives better phenomenology for $a_{1}$ decay
widths and the pion charge radius~\cite{Ko94}. This needs to be
further explored.

\section{Double-meson vertices}
\label{sec:double-meson}
\subsection{Vector double-meson vertices}
The vector double-meson vertices are obtained from an expansion
of Eq.~(\ref{phivc}). The expansion to second order in the meson
fields is given by
\begin{equation}
  \Phi_{\rm vc} =
          \frac{1}{\sqrt{2}}\,\gamma^{\mu}(\lambda\rho_{\mu})
  -\frac{i(1-2g_{V}f_{1}h)}{4\sqrt{2}g_{V}f_{1}^{2}}\,\gamma^{\mu}
      \left[(\lambda\pi),\partial_{\mu}(\lambda\pi)\right]
  -\frac{i}{2\sqrt{2}f_{1}}\,\gamma^{\mu}
      \left[(\lambda\pi),(\lambda A_{\mu})\right] + \ldots
\end{equation}
Hence, it should be obvious that the pair interaction Lagrangians
can be obtained by a simple replacement of the meson fields in the
single-meson interaction Lagrangians. For example, the two-pion
interaction Lagrangian is obtained from the vector version of
Eq.~(\ref{Lbar8}) with the replacement
\begin{equation}
    \gamma^{\mu}\bbox{\rho}_{\mu} \longrightarrow
    \gamma^{\mu}(\bbox{\pi}\times\partial_{\mu}\bbox{\pi}),
\end{equation}
which gives
\begin{eqnarray}
  m^{2}_{\pi}{\cal L}_{(\pi\pi)} &=&
    -g_{N\!N(\pi\pi)}(\overline{N}\gamma^{\mu}\bbox{\tau}N)
       \!\cdot\!(\bbox{\pi}\!\times\!\partial_{\mu}\bbox{\pi})
    -g_{\Xi\Xi(\pi\pi)}(\overline{\Xi}\gamma^{\mu}\bbox{\tau}\Xi)
       \!\cdot\!(\bbox{\pi}\!\times\!\partial_{\mu}\bbox{\pi})
                                                      \nonumber\\
  && -g_{\Lambda\Lambda(\pi\pi)}
          (\overline{\Lambda}\gamma^{\mu}\bbox{\Sigma}
          +\overline{\bbox{\Sigma}}\gamma^{\mu}\Lambda)
       \!\cdot\!(\bbox{\pi}\!\times\!\partial_{\mu}\bbox{\pi})
     +ig_{\Sigma\Sigma(\pi\pi)}(\overline{\bbox{\Sigma}}\times
        \gamma^{\mu}\bbox{\Sigma})\!\cdot\!
       (\bbox{\pi}\!\times\!\partial_{\mu}\bbox{\pi}).
\end{eqnarray}
Here we introduced the square of the charged-pion mass to make the
coupling constants dimensionless. Substituting the appropriate
renormalization factors and ${\textstyle\frac{1}{2}}g_{V}=g_{N\!N\rho}$,
the coupling constants are given by
\begin{equation}
   g_{B'B(\pi\pi)} = \frac{m_{\pi}^{2}}{4f_{1}^{2}}\,
       \frac{2Z_{\pi}-1}{Z_{\pi}}\,\frac{g_{B'B\rho}}{g_{N\!N\rho}},
\end{equation}
for $B'B=N\!N$, $\Xi\Xi$, $\Lambda\Sigma$, and $\Sigma\Sigma$. Note that
by choosing~\cite{Wei67,Wes67,Sch67} $Z_{\pi}={\textstyle\frac{1}{2}}$,
i.e., making the assumption that $m_{a_{1}}=\sqrt{2}m_{\rho}$,
all the $(\pi\pi)$ pair interactions are absent. This agrees with
the assumption of vector meson dominance which in its most stringent
form states that those multiple-meson interactions which can arise
through the exchange of one single vector meson should not also occur
directly~\cite{Wes67}.
However, experimentally $m_{a_{1}}\neq\sqrt{2}m_{\rho}$, and so here
the $(\pi\pi)$ pair interactions are still present in the interaction
Lagrangian. We can also view the $(\pi\pi)$ pair exchange as to
proceed via $\rho$ exchange where the $\rho$-meson propagator is
approximated by a constant (which should be adequate for a heavy meson
at low energies). Comparing the coupling constant for this and any other
pair vertex with the effective coupling constant as obtained in such
a meson saturation picture then gives an indication of how good this
picture really is.

The $(\pi K)$ and $(\eta_{8}K)$ combinations occur at the same place
in the $\Phi_{\rm vc}$ matrix as the $K^{\ast}$ fields. Therefore, the
$(\pi K)$ and $(\eta_{8}K)$ interactions are obtained by substituting,
respectively,
\begin{eqnarray}
    \gamma^{\mu}K^{\ast}_{\mu} &\longrightarrow&
    -i\gamma^{\mu}\bbox{\tau}\!\cdot\!(\bbox{\pi}\partial_{\mu}K
        -K\partial_{\mu}\bbox{\pi}),       \nonumber\\
    \gamma^{\mu}K^{\ast}_{\mu} &\longrightarrow& -i\gamma^{\mu}
    (\eta_{8}\partial_{\mu}K-K\partial_{\mu}\eta_{8}).
\end{eqnarray}
This gives the coupling constants
\begin{eqnarray}
   g_{B'B(\pi K)} &=& \frac{m_{\pi}^{2}}{4\sqrt{2}f_{1}^{2}}\,
      \frac{2Z_{K\pi}-1}{Z_{K\pi}}\,
      \frac{g_{B'BK^{\ast}}}{g_{N\!N\rho}},  \nonumber\\
   g_{B'B(\eta_{8}K)} &=& \frac{m_{\pi}^{2}}{4\sqrt{2}f_{1}^{2}}\,
      \frac{2Z_{K\eta_{8}}-1}{Z_{K\eta_{8}}}\,
      \frac{\sqrt{3}g_{B'BK^{\ast}}}{g_{N\!N\rho}},
\end{eqnarray}
for $B'B=\Lambda N$, $\Xi\Lambda$, $\Sigma N$, and $\Xi\Sigma$.
Here we defined averaged renormalization constants $Z^{2}_{ab}=
Z_{a}Z_{b}$ to simplify the expressions.

In the $\Phi_{\rm vc}$ matrix, the two-kaon interactions occur on the
diagonal and on the same off-diagonal places as the $\bbox{\rho}$ fields.
We can therefore split up the contributions into two parts. One part
behaves like an isoscalar octet field, and so we have the replacement
\begin{equation}
   \gamma^{\mu}\omega_{8\mu} \longrightarrow
   i\gamma^{\mu}(\overline{K}\partial_{\mu}K-
                 \overline{K_{c}}\partial_{\mu}K_{c}),
\end{equation}
and the coupling constants
\begin{equation}
   g_{BB(KK)} = \frac{m_{\pi}^{2}}{8f_{1}^{2}}\,
      \frac{2Z_{K}-1}{Z_{K}}\,
      \frac{\sqrt{3}g_{BB\omega_{8}}}{g_{N\!N\rho}}.
\end{equation}
The remaining part behaves like an isovector field ${\bf K}_{2\mu}$
which can be written as
\begin{equation}
   {\bf K}_{2\mu}=(\overline{K}\bbox{\tau}\partial_{\mu}K)
     +(\overline{K_{c}}\bbox{\tau}\partial_{\mu}K_{c}), \label{KKvecmu}
\end{equation}
The Lagrangian for this type of two-kaon interaction is obtained by
making the substitution $\bbox{\rho}_{\mu}\rightarrow i{\bf K}_{2\mu}$,
which gives the coupling constants
\begin{equation}
   g_{B'B(K\tau K)} = \frac{m_{\pi}^{2}}{8f_{1}^{2}}\,
      \frac{2Z_{K}-1}{Z_{K}}\,\frac{g_{B'B\rho}}{g_{N\!N\rho}},
\end{equation}
where again $B'B=N\!N$, $\Xi\Xi$, $\Lambda\Sigma$, and $\Sigma\Sigma$.

The double-meson interactions consisting of a pseudoscalar and
an axial-vector meson are analogously obtained by the substitutions
\begin{eqnarray}
   \gamma^{\mu}\bbox{\rho}_{\mu} &\longrightarrow&
       \gamma^{\mu}(\bbox{\pi}\!\times\!{\bf A}_{\mu}),   \nonumber\\
   \gamma^{\mu}K^{\ast}_{\mu} &\longrightarrow&
       -i\gamma^{\mu}\bbox{\tau}\!\cdot\!(\bbox{\pi}K_{1\mu}
             -K{\bf A}_{\mu}),                            \nonumber\\
   \gamma^{\mu}K^{\ast}_{\mu} &\longrightarrow& -i\gamma^{\mu}
       (\eta_{8}K_{1\mu}-Ka_{8\mu}),                      \nonumber\\
   \gamma^{\mu}\omega_{8\mu} &\longrightarrow&
       i\gamma^{\mu}(\overline{K}K_{1\mu}
             -\overline{K_{c}}K_{1\mu\,c}),               \nonumber\\
   \gamma^{\mu}\bbox{\rho}_{\mu} &\longrightarrow&
       i\gamma^{\mu}(\overline{K}\bbox{\tau}K_{1\mu}
             +\overline{K_{c}}\bbox{\tau}K_{1\mu\,c}).
\end{eqnarray}
Including averaged renormalization constants, the coupling constants
are given by
\begin{eqnarray}
  g_{B'B(\pi a_{1})} &=& \frac{m_{\pi}}{\sqrt{Z_{\pi}}f_{1}}\,
        g_{B'B\rho},      \nonumber\\
  g_{B'B(\pi K_{1})} &=&-g_{B'B(Ka_{1})}
        = \frac{m_{\pi}}{\sqrt{Z_{K\pi}}f_{1}}\,
        {\textstyle\sqrt{\frac{1}{2}}}g_{B'BK^{\ast}},   \nonumber\\
  g_{B'B(\eta_{8}K_{1})} &=&-g_{B'B(Ka_{8})}
        = \frac{m_{\pi}}{\sqrt{Z_{K\eta_{8}}}f_{1}}\,
        {\textstyle\sqrt{\frac{3}{2}}}g_{B'BK^{\ast}},   \nonumber\\
  g_{B'B(KK_{1})} &=& \frac{m_{\pi}}{2\sqrt{Z_{K}}f_{1}}\,
        \sqrt{3}g_{B'B\omega_{8}}, \nonumber\\
  g_{B'B(K\tau K_{1})} &=& \frac{m_{\pi}}{2\sqrt{Z_{K}}f_{1}}\,
        g_{B'B\rho},
\end{eqnarray}
with the relevant substitutions for the $B'B$ baryon combinations.

\subsection{Axial-vector double-meson vertices}
The expansion of Eq.~(\ref{phiax}) to second order in the meson
fields gives
\begin{equation}
  \Phi_{\rm ax} =
          \frac{1}{\sqrt{2}}\,\gamma^{5}\gamma^{\mu}(\lambda A_{\mu})
  +\frac{1-g_{V}f_{1}h}{\sqrt{2}g_{V}f_{1}}\,
          \gamma^{5}\gamma^{\mu}\partial_{\mu}(\lambda\pi)
  -\frac{i(1-g_{V}f_{1}h)}{2\sqrt{2}f_{1}}\,\gamma^{5}\gamma^{\mu}
      \left[(\lambda\pi),(\lambda\rho_{\mu})\right] + \ldots
\end{equation}
Completely analogous to the previous section, we make the substitutions
\begin{eqnarray}
   \gamma^{5}\gamma^{\mu}{\bf A}_{\mu} &\longrightarrow& \gamma^{5}
      \gamma^{\mu}(\bbox{\pi}\!\times\!\bbox{\rho}_{\mu}), \nonumber\\
   \gamma^{5}\gamma^{\mu}K_{1\mu} &\longrightarrow& -i\gamma^{5}
      \gamma^{\mu}\bbox{\tau}\!\cdot\!
      (\bbox{\pi}K^{\ast}_{\mu}-K\bbox{\rho}_{\mu}),       \nonumber\\
   \gamma^{5}\gamma^{\mu}K_{1\mu} &\longrightarrow& -i\gamma^{5}
      \gamma^{\mu}(\eta_{8}K^{\ast}_{\mu}-K\omega_{8\mu}), \nonumber\\
   \gamma^{5}\gamma^{\mu}a_{8\mu} &\longrightarrow& i\gamma^{5}
      \gamma^{\mu}(\overline{K}K^{\ast}_{\mu}
       -\overline{K_{c}}K^{\ast}_{\mu\,c}),                \nonumber\\
   \gamma^{5}\gamma^{\mu}{\bf A}_{\mu} &\longrightarrow&
       i\gamma^{5}\gamma^{\mu}(\overline{K}\bbox{\tau}K^{\ast}_{\mu}
             +\overline{K_{c}}\bbox{\tau}K^{\ast}_{\mu\,c}).
\end{eqnarray}
The coupling constants are most easily expressed in terms of
$g_{N\!N\rho}$ and the pseudovector coupling constants:
\begin{eqnarray}
   g_{B'B(\pi\rho)} &=& 2g_{N\!N\rho}f_{B'B\pi},       \nonumber\\
   g_{B'B(\pi K^{\ast})} &=& -g_{B'B(K\rho)} =
         \sqrt{2}g_{N\!N\rho}f_{B'BK},                 \nonumber\\
   g_{B'B(\eta_{8}K^{\ast})} &=& -g_{B'B(K\omega_{8})} =
         \sqrt{6}g_{N\!N\rho}f_{B'BK},                 \nonumber\\
   g_{B'B(KK^{\ast})} &=& \sqrt{3}g_{N\!N\rho}f_{B'B\eta_{8}},\nonumber\\
   g_{B'B(K\tau K^{\ast})} &=& g_{N\!N\rho}f_{B'B\pi}.
\end{eqnarray}

\subsection{Possible extensions}
\label{subsec:extension}
Given the transformation property of the $\Phi_{\rm sc}$ and
$\Phi_{\rm ax}$ fields, we can of course arbitrarily add more
interaction Lagrangians of the type (\ref{LIsu3}) with higher orders
in the meson fields. In general this will introduce a number of new
free parameters. Because there is no guarantee that we can actually
construct a baryon-baryon potential without any free parameters for
the coupling constants, it might be very useful to still have some
flexibility (i.e., free parameters) in the model.
It will be convenient, of course, if at the same time we can still
satisfy the empirical constraints for the single-meson coupling
constants as given in Sec.~\ref{sec:single-meson}. With this purpose in
mind, we therefore now discuss some possible extensions in more detail.

Although most of the scalar meson masses are already close to 1 GeV,
the low mass of $\sim$500 MeV in the two-pole approximation~\cite{Sch71}
to the broad $\varepsilon(760)$ meson makes the double-scalar
interactions worthwhile to be investigated.
We can extend the scalar interaction Lagrangian by adding
$\sqrt{2}(g_{ss}/m_{\pi})(\Phi_{\rm sc})^{2}$ to $\Phi_{\rm sc}$
of Eq.~(\ref{phisc}), where the charged-pion mass is introduced to make
the coupling constant dimensionless. Since the octet and singlet
parts in the interaction Lagrangian (\ref{LIsu3}) have independent
coupling constants, we can also define two independent scalar-scalar
coupling constants. Hence, writing $X=\lambda_{c}s_{c}$ we have the
substitution
\begin{equation}
   g^{\rm oct}_{\rm sc}\Phi^{(8)}_{\rm sc}\longrightarrow
      \frac{1}{\sqrt{2}}\left\{ g^{\rm oct}_{\rm sc}\left(F+X\right)
      +\frac{g^{(8)}_{ss}}{m_{\pi}}
      \left(F^{2}+(FX+XF)+X^{2}\right) \right\},
\end{equation}
and a similar expression for the singlet part. We still want to satisfy
the baryon-mass relations of Sec.~\ref{subsec:scalar}, which gives the
constraints
\begin{eqnarray}
  {\textstyle\sqrt{\frac{1}{6}}} \left[
           g^{\rm sin}_{\rm sc}(2f_{1}+f_{2})
          +\frac{g_{ss}^{(1)}}{m_{\pi}}(2f_{1}^{2}+f_{2}^{2}) \right]
     &=& 1153.5\ {\rm MeV},                             \nonumber\\
  \left[g^{\rm oct}_{\rm sc}(f_{2}-f_{1})+\frac{g_{ss}^{(8)}}{m_{\pi}}
         (f_{2}^{2}-f_{1}^{2})\right]   &=& 136.5\ {\rm MeV},
                                                  \label{constraint}
\end{eqnarray}
where we substituted $M_{0}=1153.5$ MeV.
Proceeding along the lines of the previous section, the evaluation
of $(\lambda_{c}s_{c})^{2}$ under the assumption of ideal mixing
allows us to read off the double-scalar interactions by making the
following substitutions
\begin{eqnarray}
   g^{\rm oct}_{\rm sc}{\bf a}_{0} &\rightarrow& g_{ss}^{(8)}\left[
      2f_{0}{\bf a}_{0}+{\textstyle\frac{1}{2}}
      (\overline{\kappa}\bbox{\tau}\kappa
      -\overline{\kappa_{c}}\bbox{\tau}\kappa_{c})\right], \nonumber\\
   g^{\rm oct}_{\rm sc}\kappa\, &\rightarrow& g_{ss}^{(8)}
      \left[(\bbox{\tau}\!\cdot\!{\bf a}_{0})\kappa
            +(f_{0}-\sqrt{2}\varepsilon)\kappa\right],    \nonumber\\
   g^{\rm oct}_{\rm sc}s_{8} &\rightarrow& g_{ss}^{(8)}
      {\textstyle\sqrt{\frac{1}{3}}}\left[{\bf a}_{0}^{2}+f_{0}^{2}
      -2\varepsilon^{2}-\overline{\kappa}\kappa\right],   \nonumber\\
   g^{\rm sin}_{\rm sc}s_{0} &\rightarrow& g_{ss}^{(1)}
      {\textstyle\sqrt{\frac{2}{3}}}\left[{\bf a}_{0}^{2}+f_{0}^{2}
      +\varepsilon^{2}+2\overline{\kappa}\kappa\right],
\end{eqnarray}
where we included the relevant coupling constants.
Note that the vacuum expectation matrix $F$ does not commute with
$\lambda_{c}s_{c}$, and so also the scalar single-meson coupling
constants need some modification. In terms of the original coupling
constants, we have
\begin{eqnarray}
   g_{B'Ba_{0}} &\longrightarrow& g_{B'Ba_{0}}
            \left[1+\frac{2f_{1}}{m_{\pi}}\,
            \frac{g_{ss}^{(8)}}{g^{\rm oct}_{\rm sc}}\right], \nonumber\\
   g_{B'B\kappa}\, &\longrightarrow& g_{B'B\kappa} \left[
            1+\frac{f_{1}+f_{2}}{m_{\pi}}\,
            \frac{g_{ss}^{(8)}}{g^{\rm oct}_{\rm sc}}\right], \nonumber\\
   g_{B'Bs_{8}} &\longrightarrow& g_{B'Bs_{8}}
            \left[1+\frac{2}{3}\,\frac{f_{1}+2f_{2}}{m_{\pi}}\,
            \frac{g_{ss}^{(8)}}{g^{\rm oct}_{\rm sc}}\right]
            +\frac{2\sqrt{2}}{3}\,\frac{f_{1}-f_{2}}{m_{\pi}}\,
             \tilde{g}_{B'Bs_{8}},                       \nonumber\\
   g_{B'Bs_{0}} &\longrightarrow& g_{B'Bs_{0}}
            \left[1+\frac{2}{3}\,\frac{2f_{1}+f_{2}}{m_{\pi}}\,
            \frac{g_{ss}^{(1)}}{g^{\rm sin}_{\rm sc}}\right]
            +\frac{2\sqrt{2}}{3}\,\frac{f_{1}-f_{2}}{m_{\pi}}\,
             \tilde{g}_{B'Bs_{0}},               \label{scalarmod}
\end{eqnarray}
where $\tilde{g}_{B'Bs_{8}}$ is obtained from the scalar version of
Eq.~(\ref{gsin}) replacing $s_{0}$ by $s_{8}$ and $g^{\rm sin}_{\rm sc}$
by $g_{ss}^{(1)}$. Similarly, $\tilde{g}_{B'Bs_{0}}$ is obtained from
the scalar version of Eq.~(\ref{goct}) replacing $s_{8}$ by $s_{0}$
and $g^{\rm oct}_{\rm sc}$ by $g_{ss}^{(8)}$.
Clearly, this type of extension allows for much more freedom in the
scalar one-boson exchanges, without the need of abandoning the
generation of the proper baryon masses. Note, however, that because of
the second constraint in Eq.~(\ref{constraint}), the resulting numerical
values for $g_{B'B\kappa}$, of some importance in $Y\!N$ potentials,
remain unaffected by these changes.

A similar example is the case where we include the matrix for the scalar
fields in the axial-vector interaction Lagrangian. This gives both
scalar-pseudoscalar and scalar--axial-vector pair interactions. Since
the axial-vector meson masses are well above 1 GeV and the scalar meson
masses are close to 1 GeV, the scalar--axial-vector exchanges are
expected to be completely negligible in baryon-baryon potential models,
and we will not consider them here.
The scalar-pseudoscalar interactions are generated by the combination
$\{\Phi_{\rm sc},\Phi_{\rm ax}\}$. In principle, another possible
combination is $i[\Phi_{\rm sc},\Phi_{\rm ax}]$, where the $i$ is
required by hermiticity. However, due to the fact that the vacuum
expectation matrix $F$ does not commute with $\Phi_{\rm ax}$, the
commutator generates a complex contribution to the baryon-baryon-kaon
coupling constants; we therefore drop this combination.
As before, by including the anticommutator combination we can introduce
two coupling constants $g_{sp}^{(8)}$ and $g_{sp}^{(1)}$, for the octet
and singlet part of the interaction Lagrangian, respectively. The form
of the various interaction Lagrangians and the expressions for the
pair coupling constants are now easy to derive. Note that the
baryon-baryon-pseudovector coupling constants are modified in an
analogous manner to Eq.~(\ref{scalarmod}).

As a final example we discuss the class of pair-meson interactions
which, within the present formalism, do not have any theoretical
constraint on the overall coupling constants.
The most simple interaction of this type contains $(\Phi_{\rm ax})^{2}$
or, equivalently, $u_{\mu}u_{\nu}$. This is also the most important one
since it contains the lightest meson (the pion).
Because the axial-vector mesons are already rather heavy, in the
following we will only consider the pseudoscalar-pseudoscalar
contributions. We have the possibility for two types of field
combinations, one symmetric and one antisymmetric in the fields:
\begin{eqnarray}
   \phi_{s} &\sim& -{\textstyle\frac{1}{2}}g^{\mu\nu}\left[
     \partial_{\mu}(\lambda\pi)\partial_{\nu}(\lambda\pi)
    +\partial_{\nu}(\lambda\pi)\partial_{\mu}(\lambda\pi)\right],
                                              \nonumber\\
   \phi_{a} &\sim& +{\textstyle\frac{i}{2}}\sigma^{\mu\nu}\left[
     \partial_{\mu}(\lambda\pi)\partial_{\nu}(\lambda\pi)
    -\partial_{\nu}(\lambda\pi)\partial_{\mu}(\lambda\pi)\right].
\end{eqnarray}
It will be convenient to identify the two-pseudoscalar contributions
with a matrix of scalar fields as in Eq.~(\ref{Smat}). For the
symmetric combination this implies the following substitutions
\begin{eqnarray}
  {\bf a}_{0} \rightarrow -{\textstyle\frac{1}{2}}g^{\mu\nu} && \left[
    {\textstyle\sqrt{\frac{1}{3}}}(\partial_{\mu}\bbox{\pi}
   \partial_{\nu}\eta_{8}+\partial_{\nu}\bbox{\pi}\partial_{\mu}\eta_{8})
   +{\textstyle\sqrt{\frac{2}{3}}}(\partial_{\mu}\bbox{\pi}
   \partial_{\nu}\eta_{0}+\partial_{\nu}\bbox{\pi}\partial_{\mu}\eta_{0})
   +{\textstyle\frac{1}{2}}(\partial_{\mu}\overline{K}\bbox{\tau}
   \partial_{\nu}K+\partial_{\nu}\overline{K}\bbox{\tau}\partial_{\mu}K)
   \right],                                      \nonumber\\
  \kappa \rightarrow -{\textstyle\frac{1}{2}}g^{\mu\nu} && \left[
   {\textstyle\sqrt{\frac{1}{2}}}\bbox{\tau}\!\cdot\!(\partial_{\mu}
    \bbox{\pi}\partial_{\nu}K+\partial_{\nu}\bbox{\pi}\partial_{\mu}K)
    -{\textstyle\sqrt{\frac{1}{6}}}
    (\partial_{\mu}\eta_{8}\partial_{\nu}K+\partial_{\nu}\eta_{8}
     \partial_{\mu}K)+{\textstyle\sqrt{\frac{4}{3}}}(\partial_{\mu}
     \eta_{0}\partial_{\nu}K+\partial_{\nu}\eta_{0}\partial_{\mu}K)
   \right],                                      \nonumber\\
  s_{8} \rightarrow -{\textstyle\frac{1}{2}}g^{\mu\nu} && \left[
   {\textstyle\sqrt{\frac{1}{3}}}\partial_{\mu}\bbox{\pi}\!\cdot\!
    \partial_{\nu}\bbox{\pi}-{\textstyle\sqrt{\frac{1}{3}}}\partial_{\mu}
    \eta_{8}\partial_{\nu}\eta_{8}+{\textstyle\sqrt{\frac{2}{3}}}
    (\partial_{\mu}\eta_{8}\partial_{\nu}\eta_{0}+\partial_{\nu}\eta_{8}
     \partial_{\mu}\eta_{0})-{\textstyle\sqrt{\frac{1}{12}}}
    (\partial_{\mu}\overline{K}\partial_{\nu}K+\partial_{\nu}
     \overline{K}\partial_{\mu}K)\right],     \nonumber\\
  s_{0} \rightarrow -{\textstyle\frac{1}{2}}g^{\mu\nu} &&
    {\textstyle\sqrt{\frac{2}{3}}}\left[\partial_{\mu}\bbox{\pi}\!\cdot\!
     \partial_{\nu}\bbox{\pi}+\partial_{\mu}\eta_{8}\partial_{\nu}\eta_{8}
    +\partial_{\mu}\eta_{0}\partial_{\nu}\eta_{0}+
    (\partial_{\mu}\overline{K}\partial_{\nu}K+\partial_{\nu}
     \overline{K}\partial_{\mu}K)\right],
\end{eqnarray}
while for the antisymmetric combinations we have
\begin{eqnarray}
  {\bf a}_{0} \rightarrow
               +{\textstyle\frac{i}{2}}\sigma^{\mu\nu} && \left[
    i\partial_{\mu}\bbox{\pi}\times\partial_{\nu}\bbox{\pi}-
    {\textstyle\frac{1}{2}}(\partial_{\mu}\overline{K}\bbox{\tau}
   \partial_{\nu}K-\partial_{\nu}\overline{K}\bbox{\tau}\partial_{\mu}K)
   \right],   \nonumber\\
  \kappa \rightarrow
               +{\textstyle\frac{i}{2}}\sigma^{\mu\nu} && \left[
   {\textstyle\sqrt{\frac{1}{2}}}\bbox{\tau}\!\cdot\!(\partial_{\mu}
    \bbox{\pi}\partial_{\nu}K-\partial_{\nu}\bbox{\pi}\partial_{\mu}K)
    +{\textstyle\sqrt{\frac{3}{2}}}
    (\partial_{\mu}\eta_{8}\partial_{\nu}K-\partial_{\nu}\eta_{8}
     \partial_{\mu}K) \right],  \nonumber\\
  s_{8} \rightarrow
             +{\textstyle\frac{i}{2}}\sigma^{\mu\nu} && \left[
     -{\textstyle\sqrt{\frac{3}{4}}}(\partial_{\mu}\overline{K}
     \partial_{\nu}K-\partial_{\nu}\overline{K}\partial_{\mu}K)\right].
\end{eqnarray}
These matrices can be substituted in the interaction Lagrangian
(\ref{LIsu3}), which contains the free parameters $g^{(8)}_{sym}$,
$g^{(1)}_{sym}$, and $\alpha_{sym}$ for the symmetric case, and
$g^{(8)}_{asym}$ and $\alpha_{asym}$ for the antisymmetric case.
Including the renormalization factors it is now straightforward to
find the pair coupling constants for each of the two-pseudoscalar
contributions, expressed in terms of these free parameters.

\section{Application to $\protect\bbox{N\!N}$}
\label{sec:application}
As a first application of the chiral-symmetry constraints given in
this paper, we like to investigate whether with the values for the
coupling constants as given in the previous sections it is indeed
possible to construct a baryon-baryon potential model which gives a
satisfactory description of the baryon-baryon scattering data.
Of course, the imposed constraints need not all be exactly true.
For example, the vector-dominance assumption that $\kappa_{\rho}=3.71$
and $\kappa_{\omega}+\kappa_{\phi}=-0.12$ is only true if these mesons
fully saturate the nucleon electromagnetic form factors. The presence of
heavier vector-meson nonets likely changes these relations~\cite{Dub91}.
Also, the SU(3) relations in Sec.~\ref{sec:single-meson} need not be
true in an exact sense. This is already clear from the fact that we have
to introduce symmetry-violating terms to generate the empirical meson
masses. The existence of a scalar meson nonet and its quark content is
still under debate, and so the assumption of an SU(3) symmetry for the
scalar mesons might even be incorrect.
On the other hand, relaxing too many constraints introduces too many
free parameters -- something we would like to avoid.
Therefore, at this stage we choose to impose {\it all\/} the constraints
and only show that the resulting $N\!N$ potential model then already
gives a very reasonable description of the scattering data.

The experience with $N\!N$ potential models that have appeared in
the literature suggests that a fully constrained potential model of
the one-boson-exchange type is unlikely to succeed~\cite{Sto95}.
On the other hand, we have already demonstrated~\cite{Rij96a,Rij96b}
that by including two-meson-exchange contributions a major improvement
in the description of the $N\!N$ scattering data can be obtained.
In order to arrive at a model which at least gives a reasonable
description of the scattering data, we found that we had to include the
double-scalar and double-pseudoscalar extensions as outlined in
Sec.~\ref{subsec:extension}. The single-meson coupling constants satisfy
the empirical constraints as discussed in Sec.~\ref{sec:single-meson}
and the pair-meson coupling constants satisfy the relations as given in
Sec.~\ref{sec:double-meson}.
The one-boson-exchange part of the potential is standard but includes
the diffractive contribution~\cite{Nag78}, while the two-meson part
can be found in, or easily derived from, Refs.~\cite{Rij96a,Rij96b}.
The potential is regularized with exponential form factors, one for
each type of meson (scalar, pseudoscalar, or vector). The single-meson
coupling constants and the exponential cutoffs for each type are given
in Table~\ref{copsing}. The pair-meson coupling constants are given in
Table~\ref{coppair}. Note that we only include meson pairs with a total
mass below $\sim$1 GeV. However, since the $\eta\eta$-exchange
contributions did not significantly improve the fit, we decided not to
include them at this stage.

The 12 free parameters of the model ($\Lambda_{S}$, $\Lambda_{P}$,
$\Lambda_{V}$, $g_{A_{2}}$, $g_{P}$, $g_{ss}^{(8)}$, $g_{ss}^{(1)}$,
$g_{sym}^{(8)}$, $g_{sym}^{(1)}$, $\alpha_{sym}$, $g_{asym}^{(8)}$, and
$\alpha_{asym}$) were determined in a fit to the Nijmegen
representation~\cite{Sto93c} of the $\chi^{2}$ hypersurface of the
$N\!N$ scattering data below $T_{\rm lab}=350$ MeV, updated with the
inclusion of new data which have been published since then.
The effective diffractive mass, $m_{P}=310$ MeV, was fixed at the
(rounded-off) value as used in the old Nijm78 potential~\cite{Nag78}.
The resulting $\chi^{2}/N_{\rm data}$ for each of the ten energy bins
is shown in Table~\ref{chi2}, in comparison with the (updated)
Nijmegen partial-wave analysis.

The $\chi^{2}/N_{\rm data}=1.75$ for the 0--350 MeV energy interval
actually compares very favorably to other potential models that
have appeared in the literature~\cite{Sto95}. As a matter of fact,
it should be realized that in this model {\it all\/} coupling constants
satisfy constraints as imposed by chiral symmetry, or empirical
constraints as discussed in Sec.~\ref{sec:single-meson}. This in
contrast to any other model that has appeared in the literature.
The model even gives a much better description of the data below
300 MeV ($\chi^{2}/N_{\rm data}=1.36$), whereas it rapidly worsens at
higher energies.
This sudden rise is probably due to the nonadiabatic expansion in the
two-meson contributions~\cite{Rij96a,Rij96b} which, strictly speaking,
is only valid below the pion production threshold ($T_{\rm lab}\approx
280$ MeV). The nonadiabatic expansion is an artifact of us working in
coordinate space, and the sudden rise in $\chi^{2}$ at higher energies
will be further investigated when we have developed a momentum-space
version where we can retain the full energy dependence in the
propagators\footnote{Alternatively, we can decide that the model should
only be used up to $T_{\rm lab}\approx300$ MeV, say. Without pursuing
this option any further, a quick and not too thorough refit already shows
that a $\chi^{2}/N_{\rm data}=1.3$ seems easily feasible.}.
At this stage we prefer to work in coordinate space for several reasons.
First, the (already rather time consuming) fit in coordinate space is
much faster than a fit in momentum space.
Second, here we only wanted to investigate whether it is indeed possible
to construct a potential model which incorporates all constraints and
still gives a satisfactory description of the data. Modifications such
as keeping the full energy dependence in the propagators rather than
making a nonadiabatic expansion can be investigated at a later stage.
Third, ultimately it is our goal to produce a high-quality potential
model which is exactly equivalent in both coordinate space and momentum
space. This allows equivalent applications in both spaces, but requires
certain approximations to be made in order to arrive at analytical
expressions in coordinate space.

We should point out that, although the number of free parameters is
not too large, the fully constrained fit is far from trivial. Due to
the constraints, the relation between the change in a parameter and
the corresponding change in the phase shifts is highly nonlinear.
It turned out to be impossible to fit all the free parameters at the
same time, and so we had to do numerous fit cycles where we fit only
an (arbitrary) subset of the parameters.
This makes it very complicated and time consuming (but not impossible)
to arrive at a satisfactory fit, and we cannot guarantee that the fit
presented here is the most optimal one. However, the present result
already clearly illustrates our main objective, viz.\ to try to construct
a potential model which gives a reasonable description of the scattering
data, and at the same time incorporates a number of empirical and
chiral-symmetry constraints. Improvements by investigating different
parameter sets, adding more parameters, or relaxing some of the
constraints will be left for the future.

\section{Summary}
We have constructed a chiral-invariant Lagrangian for the meson-baryon
sector, where the mesons consist of the nonets of scalar, pseudoscalar,
vector, and axial-vector mesons, and the baryons are the members of
the baryon octet. Although we mainly focussed on the meson-baryon
interaction Lagrangian, we briefly indicated what the Lagrangian in
the meson sector looks like, and how it can be made to reproduce
the correct phenomenology by allowing for some small contributions
which violate the chiral symmetry.

In the meson-baryon sector, the chiral symmetry imposes constraints
on the coupling constants of the meson-pair vertices. This allows
us to express all meson-pair coupling constants in terms of the
single-meson vertex coupling constants. These single-meson vertex
coupling constants, in principle, can all be fixed by using
experimental data such as meson masses, baryon masses, meson decay
parameters, and meson mixing angles.

As a first step in the application of the meson-baryon Lagrangian, we
demonstrate that it is possible to construct an $N\!N$ potential which
has all the theoretical and empirical constraints as discussed in
Secs.~\ref{sec:single-meson} and \ref{sec:double-meson}. The model
gives a very satisfactory description of the $N\!N$ scattering data.
It even has a slightly better quality than other (often more
phenomenological) potential models that have appeared in the literature,
where this is the first model where such strict constraints on the
coupling constants have successfully been imposed. We should mention
that this result could only be achieved after the inclusion of the
two-meson contributions. Improvements and an extension to the $Y\!N$
sector are presently under investigation.
The success of the present model gives us hope that it might indeed be
possible to ultimately arrive at a potential model which not only
provides a high-quality description of the scattering data, but which
is also consistent with the symmetries of QCD.

\acknowledgments
Discussions with H.\ Fearing, N.\ Mobed, and D.\ Phillips are
gratefully appreciated. This work was supported by the Natural
Sciences and Engineering Research Council of Canada.

\appendix
\section*{}
We follow Ref.~\cite{Bar69} and show that it is possible to find a
nonlinear transformation that transforms away the octet of original
pseudoscalar fields, while leaving the scalar vacuum expectation
matrix $F$ invariant. We write the scalar nonet as
$\lambda_{c}\sigma_{c}=F+\Sigma$, where the diagonal matrix
$F=(f_{1},f_{2},f_{3})$ is the nonvanishing vacuum expectation matrix
of the scalar fields. The parameters $f_{i}$ can be anything, but
in the following we assume that $f_{i}+f_{j}\neq0$ for all $\{i,j\}$.
The pseudoscalar octet is written as $\Pi=\lambda_{a}\pi_{a}$.
We want to find a traceless matrix $P$ (to be identified as an octet
of new pseudoscalar fields), such that
\begin{equation}
   F+\Sigma+i\Pi = e^{iP}(F+X+iY)e^{iP},
\end{equation}
where $X$ and $Y$ are sets of different scalar and pseudoscalar
fields, respectively. For that purpose, we formally write~\cite{Bar69}
\begin{equation}
   P=\alpha\sum_{k=0}^{\infty}\alpha^{k}P_{k}, \ \ \
   X=\sum_{k=0}^{\infty}\alpha^{k}X_{k}, \ \ \
   Y=\sum_{k=0}^{\infty}\alpha^{k}Y_{k},
\end{equation}
and try to solve for
\begin{equation}
   F/\alpha+\Sigma+i\Pi=e^{iP}(F/\alpha+X+iY)e^{iP},
\end{equation}
in each order of $\alpha$. To order $\alpha^{-1}$ we have an
identity. To order $\alpha^{0}$ we can write
\begin{eqnarray}
  &&  X_{0} = \Sigma, \ \ \ \
  Y_{0}=\frac{\Pi_{11}f_{2}f_{3}+\Pi_{22}f_{1}f_{3}+\Pi_{33}f_{1}f_{2}}
        {f_{2}f_{3}+f_{1}f_{3}+f_{1}f_{2}}\,\openone_{3}, \nonumber\\
  && (P_{0})_{ij} = (f_{i}+f_{j})^{-1}\left[\Pi-Y_{0}\right]_{ij},
\end{eqnarray}
where we chose $Y_{0}$ such that ${\rm Tr}(P)=0$. This is not the
only possible choice for $Y_{0}$, but it will turn out to be a very
convenient one. (Note that in Ref.~\cite{Bar69} a different choice is
made, but there $P_{0}$ is not traceless, contrary to what is claimed.)
This procedure can be extended to all orders in $\alpha$, and we
always find that
\begin{equation}
    Y_{n}+\{F,P_{n}\}=g_{n},
\end{equation}
with $g_{n}$ a nonlinear function of $P_{k}$, $X_{k}$, and $Y_{k}$
with $k=0,\ldots,n-1$. Hence, in analogy to the $n=0$ case, we can
always define a traceless matrix $P_{n}$ and at the same time keep
$Y_{n}$ proportional to $\openone_{3}$. Substituting $u=\exp(iP)$,
the definition (\ref{chipm}) gives
\begin{equation}
   \chi_{+} = F+X, \ \ \ \ \chi_{-} = iY,
\end{equation}
where $X$ represents the new scalar octet and where $Y$ (because
it is proportional to $\openone_{3}$) can be identified with a new
isosinglet pseudoscalar field, not present before. Clearly,
different choices for $Y_{n}$, not proportional to $\openone_{3}$,
will introduce unwanted contributions to the isovector ($\pi$) and/or
isoscalar ($\eta_{8}$) pseudoscalar fields.

\begin{table}
\caption{Single-meson $N\!N$ coupling constants and exponential
         cutoff masses for the constrained $N\!N$ model. The single-meson
         coupling constants are divided by $\sqrt{4\pi}$.}
\begin{tabular}{cccccccccccccc}
  \multicolumn{3}{c}{$g$-scalar} & \multicolumn{3}{c}{$f$-pseudosc.}
& \multicolumn{3}{c}{$g$-vector} & \multicolumn{3}{c}{$f$-vector}
& \multicolumn{2}{c}{$g$-diffr.} \\
 $a_{0}$ & $\varepsilon$ & $f_{0}$ & $\pi$ & $\eta$ & $\eta'$ & $\rho$
 & $\omega$ & $\phi$ & $\rho$ & $\omega$ & $\phi$ & $A_{2}$ & $P$ \\
\tableline
  2.599 & 3.273 & --0.978 & 0.270 & 0.117 & 0.075 &
  0.711 & 2.408 & --0.183 & 2.634 & 0.297 & 0.045 & 2.497 & 2.126 \\
  \multicolumn{3}{c}{$\Lambda_{S}=590.0$ MeV} &
  \multicolumn{3}{c}{$\Lambda_{P}=870.0$ MeV} &
  \multicolumn{3}{c}{$\Lambda_{V}=845.0$ MeV} & & & & &
\end{tabular}
\label{copsing}
\end{table}

\begin{table}
\caption{Pair-meson coupling constants for the constrained $N\!N$ model.
         The coupling constants are divided by $4\pi$.
         For the double-pseudoscalar $\pi\pi$ and $KK$ contributions
         we distinguish interactions symmetric and antisymmetric
         in the fields.}
\begin{tabular}{ccccccccccccc}
 & & & & & & & \multicolumn{3}{c}{symmetric}
             & \multicolumn{3}{c}{antisymmetric} \\
 $\pi\pi$ & $KK$ & $K\tau K$ & $\pi\rho$ & $\varepsilon\varepsilon$ &
 $\pi\eta$ & $\pi\eta'$ & $\pi\pi$ & $KK$ & $K\tau K$ &
 $\pi\pi$ & $KK$ & $K\tau K$ \\
\tableline
  --0.030 & --0.028 & --0.009 &   0.384 & 2.463 & 0.007 & 0.004 &
    0.045 &   0.007 &   0.005 & --0.050 & 0.082 & 0.033
\end{tabular}
\label{coppair}
\end{table}

\narrowtext
\begin{table}
\caption{$\chi^{2}$ and $\chi^{2}$ per datum ($\chi^{2}_{\rm p.d.p.}$)
         at the 10 energy bins for the updated partial-wave analysis
         (PWA) and the constrained $N\!N$ potential. $N_{\rm data}$
         lists the number of data within each energy bin. The bottom
         line gives the results for the total 0--350 MeV interval.}
\begin{tabular}{r@{--}lrrrrr}
    \multicolumn{2}{c}{} &  & \multicolumn{2}{c}{PWA}
  & \multicolumn{2}{c}{potential} \\
  \multicolumn{2}{c}{Bin(MeV)} & $N_{\rm data}$ & $\chi^{2}$
     & $\chi^{2}_{\rm p.d.p.}$ & $\chi^{2}$ & $\chi^{2}_{\rm p.d.p.}$ \\
\tableline
 0.0&0.5 &  145 &  144.45 & 0.996 &  155.73 & 1.07 \\
 0.5&2   &   68 &   43.08 & 0.633 &   56.48 & 0.83 \\
   2&8   &  110 &  105.11 & 0.956 &  157.41 & 1.43 \\
   8&17  &  294 &  275.24 & 0.936 &  335.07 & 1.14 \\
  17&35  &  359 &  294.58 & 0.821 &  387.39 & 1.08 \\
  35&75  &  585 &  565.30 & 0.966 & 1045.22 & 1.79 \\
  75&125 &  399 &  409.62 & 1.027 &  443.18 & 1.11 \\
 125&183 &  760 &  823.82 & 1.084 & 1249.44 & 1.64 \\
 183&290 & 1046 & 1022.72 & 0.978 & 1306.90 & 1.25 \\
 290&350 &  992 &  993.49 & 1.002 & 3197.26 & 3.22 \\
\tableline
   0&350 & 4758 & 4677.42 & 0.983 & 8334.09 & 1.75
\end{tabular}
\label{chi2}
\end{table}

\end{document}